\documentclass[10pt]{article}
\usepackage{mathpazo}
\usepackage{latexsym}
\usepackage{a4wide}
\usepackage{amsmath}
\usepackage{amssymb}
\usepackage{tikz}
\usetikzlibrary{calc}
\usepackage{enumitem}
\usepackage{multirow}
\usepackage{version}
\usepackage{hyperref}
\usepackage{subcaption}
\usepackage{graphicx}
\usepackage{comment}
\usepackage[textwidth=30mm]{todonotes}

\newtheorem{thm}{Theorem}[section]

\newtheorem{cor}[thm]{Corollary}

\newtheorem{exl}[thm]{Example}

\newtheorem{lemma}[thm]{Lemma}

\newcommand{\N}{\mathbb{N}}
\newcommand{\PP}{\mathbb{P}}

\newcommand{\I}{{\rm I}}
\newcommand{\II}{{\rm II}}

\numberwithin{equation}{section}
\newenvironment{proof}{\vspace{-0.1cm}\noindent\textsc{Proof:}}{$\Box$\medskip}

\begin{document}
\tikzstyle{state}=[circle,thick,draw=black!80]
\tikzstyle{term}=[rectangle,thick,draw=black!80]
\tikzstyle{andsoon}=[rectangle,thick,draw=white!80,fill=white!20]
\tikzstyle{andsoonfill}=[rectangle,rounded corners, text width=8em, text centered,thick,draw=black!80,fill=black!10]
\tikzstyle{nameex}=[rectangle,thick]

\title{Random perfect information games\footnote{We are grateful to 
Xavier Venel for a fruitful discussion.}}
\author{J\'{a}nos Flesch\footnote{Department of Quantitative Economics, Maastricht University, P.O.Box 616, 6200 MD, The Netherlands. E-mail: j.flesch@maastrichtuniversity.nl.}\ \and Arkadi Predtetchinski\footnote{Department of Economics, Maastricht University, P.O.Box 616, 6200 MD, The Netherlands. E-mail: a.predtetchinski@maastrichtuniversity.nl.} \and Ville Suomala\footnote{Department of Mathematical sciences, University of Oulu, P.O.Box 8000, FI--90014, Finland. E-mail: \texttt{ville.suomala@oulu.fi}}}

\maketitle

\begin{abstract}
\noindent The paper proposes a natural measure space of zero-sum perfect information games with upper semicontinuous payoffs. Each game is specified by the game tree, and by the assignment of the active player and of the capacity to each node of the tree. The payoff in a game is defined as the infimum of the capacity over the nodes that have been visited during the play. The active player, the number of children, and the capacity are drawn from a given joint distribution independently across the nodes. We characterize the cumulative distribution function of the value $v$ using the fixed points of the so-called value generating function. The characterization leads to a necessary and sufficient condition for the event $v \geq k$ to occur with positive probability. We also study probabilistic properties of the set of Player I's $k$-optimal strategies and the corresponding plays.
\end{abstract}

\noindent\textbf{Keywords:} Zero-sum game, perfect information, value, Galton-Watson measure, branching process.\bigskip

\noindent\textbf{MSC2020 classification codes:} Primary: 91A70
(Spaces of games), 91A25 (Dynamic games), 91A18 (Games in extensive form). Secondary:
Applications of branching processes (60J85).\bigskip

\tableofcontents

\section{Introduction}
Perfect information games are probably one of the most thoroughly researched classes of games. Descriptive set theory (Martin \cite{Martin75}, Moschovakis \cite{Moschovakis}), computer science (Apt and Gr\"{a}del \cite{AptGradel12}), logic (Van Benthem \cite{Benthem14}), economics (e.g. Harris \cite{Harris}) all employ their own distinct methodologies to study perfect information games. Among these, two-player zero-sum games with semicontinuous payoff function are arguably the simplest kind, intimately related to games with closed winning sets introduced by Gale and Stewart \cite{GaleStewart}. It is this, relatively simple class of infinite perfect information games that our study is devoted to.

In this paper we wish to take a probabilistic point of view on perfect information games. In a nutshell, this view amounts to the following: we consider a particular space of games and a natural probability measure over this space. Nature selects a game randomly according to the given measure, the two players observe the entire realization of the game, and the play commences. It is important to stress at the outset that, once the game has been chosen by nature, the players face no further randomness: they observe the realization of the game before the play begins, and may adjust their strategies depending on the game at hand. The most general question we are interested in is what the distribution of value is.

The probabilistic point of view described above is certainly no novelty in game theory. One-shot games with randomly generated payoffs have been scrutinized at least since Goldman's study of the probability for a randomly drawn matrix game to have a saddle point (\cite{Goldman}). Researchers examined the probability that a game possesses a pure Nash equilibrium, the distribution of the number of pure Nash equilibria, as well as questions related to convergence of learning processes. Rather than reviewing an extensive literature on random one-shot games, we refer to the two recent studies (Amiet, Collevecchio, Scarsini, and Zhong \cite{Scarsini} and Pei and Takahashi \cite{PeiTakahashi}) that contain a survey of the literature. 

On the other hand, randomly generated extensive form games have received less attention. One notable exception is Arieli and Babichenko \cite{Arieli} who consider extensive form perfect information games of given length where the payoffs at end nodes are randomly generated. Their focus is on the asymptotic distribution of subgame perfect equilibrium payoffs as the length of the games increases. 

Recently, percolation theory produced works on random perfect information games, as researchers considered adversarial versions of the classical percolation problem. Basu, Holroyd, Martin, and Wästlund \cite{Basu} examine zero-sum perfect information games played on a random graph obtained from $\mathbb{Z}^d$ by removing each node with a certain probability. The players take turns to move a token along the edges of the graph. Each node of the graph is only allowed to be visited by the token once. A player who has no legal moves is declared a loser. Holroyd, Marcovici, and Martin \cite{HolroydProb} consider a game on the square lattice $\mathbb{Z}^2$, whose every node is randomly classified as either a ``trap", or a ``target", or an ``open" node. A player can either move the token to the node on the right or to the node above of its current position. If a player moves a token into a target (respectively, a trap), she receives a payoff of $+1$ (respectively, $-1$), and if no traps or targets are ever visited, both players receive the payoff $0$. The study most closely related to ours is Holroyd and Martin \cite{Holroyd}. There three classical combinatorial games (normal, mis\`{e}re, and escape games) are played on a tree drawn randomly from a Galton-Watson measure.\medskip

\noindent\textsc{The model:} Two players, I (Alice) and II (Bob), play against each other in a perfect information game that is chosen randomly by nature. Nature chooses the game tree, selects the active player (either I or II) for each node of the tree, and assigns to every node a non-negative number, called a capacity. The number of children (i.e. the successors) of a node in the tree, the active player, and the capacity are being drawn by nature from a joint distribution independently across the nodes. For a given node, however, the active player, the capacity, and the number of children might well be interdependent. Formally, a measure space of games is a space of marked trees (as has been introduced by Neveu \cite{Neveu86}) and is a natural generalization of the Galton-Watson measure space of trees. 

Player I's payoff in a game is defined as the infimum of the capacity encountered in the course of play. Note that in any realization of the game, the game tree is finitely branching, and Player I's objective function is bounded and upper semicontinuous.

Before play commences, both players observe the entire game, including the tree, and the assignment of the active players and of the capacity to its nodes. They are free to choose a strategy depending on the realization of the game at hand. Each realization of the game has a value, Player I has an optimal pure strategy, and Player II a pure $\epsilon$-optimal strategy for every error term $\varepsilon>0$. 

The model shares several features with, and complements those in Arieli and Babichenko \cite{Arieli}, Holroyd, Marcovici, and Martin \cite{HolroydProb}, and Holroyd and Martin \cite{Holroyd}.  

As in Holroyd and Martin \cite{Holroyd} the game tree in our model is drawn from a Galton-Watson measure.

In a departure from Holroyd and Martin \cite{Holroyd} we employ a random assignment of players to the nodes of the game tree, an idea we borrow from Arieli and Babichenko \cite{Arieli}. The probability for Player I to be assigned to a node of the game tree (called I's activation probability) plays a key role in the analysis; in particular, it is driving some of the interesting phase transitions in the model. The non-adversarial case when Player I is assigned as an active player with probability 1 serves as a natural benchmark. The main advantage of the random assignment of an active player is that this modelling choice leads to a relatively tractable characterization of the distribution of the value. 

A novel feature of the model is the nature of the payoff function. In Holroyd, Marcovici and Martin \cite{HolroydProb} (as well as in the two antecedent works Basu, Holroyd, Martin, W\"{a}stlund \cite{Basu}, and Holroyd and Martin \cite{Holroyd}) all infinite branches are assigned the same payoff. This is but  one of the specifications our model allows for. The possibility of choosing a joint distribution of the active player, the capacity, and the number of offspring affords our model a lot of flexibility and allows it to accommodate many examples of non-trivial payoff functions. Some of the examples are these.

If a childless node has zero capacity, while a node with at least one child a capacity $1$, we obtain a payoff function assigning the payoff $0$ to all the end nodes and the payoff $1$ to all the infinite branches. If the capacity equals the number of offspring, we obtain a payoff function that equals the least number of children over the nodes visited in the course of the play. One other specification of interest is obtained if all nodes have $2$ children, and the capacity is uniformly distributed on $[0,1]$. This yields a game on a complete binary tree, equipped with a non-trivial payoff function over the Cantor space of plays. 

The model also accommodates certain classes of finite games. Indeed, if the mean of the offspring distribution is smaller than one, then the game tree is finite almost surely. If we specify the capacity to be $1$ whenever a node has at least one child, and to be uniformly distributed on $[0,1]$ if a node has no children, we obtain a finite game with the payoffs independently assigned to the end nodes of the tree.

Two examples are developed in detail in the main body of the paper. We also attend closely to a special case of the model called an escape model. In the escape model the end nodes, i.e. nodes without children, have capacity equal to zero. Under this specification, Player I's primary objective is to avoid the end nodes of the game tree (and thus ``escape to infinity"), for reaching such a node results in the lowest possible payoff, zero. 
\medskip 

\noindent\textsc{The results:} We split the results into three groups: I. the main results on the fixed-point characterization of the distribution of the value, II. corollaries on the properties of the value's distribution, and III. applications.\medskip  

\noindent\textsc{I. The main result:} The main result of the paper is a fixed point characterization of the distribution of the value: given $k > 0$, the probability that the value $v$ is strictly less than $k$ is shown to be a fixed point of the so-called value generating function (\emph{vgf} for short). The key consequence of the result is a criterion for the probability of the value being at least $k$ to be positive. This criterion generalizes the classical condition for (non)-extinction of a branching process. 

To obtain the result, we rely on a familiar technique of truncating a game at some period $t$, and use the fact that the value of the truncated games converges to the value of the original game as the truncation horizon increases. The role of the value generating function in our analysis is somewhat similar to the role of the  probability generating function in the study of branching processes. Intuitively, the vgf maps the distribution of the value at the next period to that at the current period. In particular, the $t$th iterate of the vgf describes the distribution of the value in the $t$-truncated game.\medskip 

\noindent\textsc{II. Corollaries:} We point out several features of the distribution of the value.\smallskip

\noindent\textit{The essential supremum of the value:} The essential supremum of the value is the highest payoff that Player I can guarantee with positive probability. We obtain an expression of the essential supremum of the value in terms of the primitives of the model (i.e. the distribution of the active players, the capacity, and the number of offspring).\smallskip  

\noindent\textit{Phase transitions in the escape model:} In the escape model, the probability of the event $v \geq k$ undergoes an interesting phase transition as a function of Player I's activation probability: the probability of the event $v \geq k$ remains zero unless Player I's activation probability is above the so-called $k$-critical level: if Player I controls too few nodes, she has no chance of obtaining a payoff greater than $k$.\smallskip

\noindent\textit{Distribution of the value conditional on the active player:} One statistic of interest is the distribution of the value conditional on the active player, or, more precisely, conditional on the event that the root of the game tree is assigned to Player I (Player II). We find (under mild and natural conditions) that the the distribution of the value at Player I's nodes first order stochastically dominates that at Player II's nodes. We also explore the relationship between the conditional probability of the event $v \geq k$ at Player I's (Player II's) nodes, as Player I's activation probability approaches its $k$-critical value.\smallskip

\noindent\textit{Asymptotic result for games on complete $n$-ary trees:} Suppose that the game tree is a complete $n$-ary tree, that is to say, each node of the tree has exactly $n$ children. How does the probability of the event $v < k$ behave as $n$ becomes large? We show that it is eventually monotone, and compute its limit as $n$ goes to infinity.\smallskip

\noindent\textit{Continuity properties of the distribution of the value:} We first look at atoms in the distribution of the value. We then turn to treat the distribution of the value as a function of the primitive distribution, i.e. of the joint distribution of the active player, the capacity, and the number of children of a node, and study under what conditions it is continuous at a particular primitive distribution.\medskip

\noindent\textsc{III. Applications:} We discuss two applications.\smallskip

\noindent\textit{Conditional game:} Starting from a game having a value of at least $k$, consider a subtree of the game tree consisting of the nodes with the value (of the corresponding subgame) of at least $k$. Restricting the original game to the subtree results in the so-called ($k$-)conditional game. 

Our motivation for the conditional game is twofold. The first is that the tree of the conditional game characterizes Player I's $k$-optimal strategies, i.e. the strategies that secure a payoff of at least $k$ no matter how Player II plays: all Player I needs to do to play $k$-optimally is never make a move leading outside the tree of the $k$-conditional game.

The second motivation is that the conditional game is a generalization of the reduced family tree of a branching process (e.g. Lyons and Peres \cite[\S 5.7]{LyonsPeres}); recall that the latter is a tree consisting of individuals with an infinite line of descent. Our findings could be seen as a game-theoretic counterpart to the textbook results on the decomposition of a branching process. 

The main result on the conditional game concerns its distribution. On the event that the original game has a value of at least $k$, the conditional game is distributed like a random game of perfect information (in the sense made precise below). We study the corresponding primitive distribution and explore some of its features. 

The conditional game is a rich source of non-trivial examples. The key feature of these examples is that the  distribution of the number of offspring at Player I's nodes is generally different from that of Player II's nodes, even if they are same in the original game.\smallskip

\noindent\textit{Avoiding Player II's nodes:} We consider a scenario where Player I is prohibited from visiting the nodes controlled by Player II, with the exception of the nodes having no children or a single child. We think of this scenario as a proxy for the situation where Player I is reluctant to concede a turn to her opponent because of e.g. security concerns. Under this scenario Player II is being deprived of any real choice in the game and is merely a dummy.

Obviously, the additional restriction makes it harder for Player I to obtain a good payoff. Somewhat surprisingly, it turns out that if Player I is able to secure a payoff of at least $k$ with positive probability in the absence of any restrictions, then she is able to do so while avoiding the nodes controlled by Player II having more than one child.\medskip

The paper is organized as follows. Section \ref{secn.model} introduces the model. Section \ref{secn.examples} illustrates the model and gives a flavour of our results by means of two examples. We keep going back to these two examples throughout the paper. Section \ref{secn.distribution} develops the key result of the paper: the characterization of the distribution of the value. Section \ref{secn.cor} presents the corollaries. Sections \ref{secn.conditional} and \ref{secn.avoid} are devoted to the two applications. The final section contains further examples, some discussion, and open questions. 

\begin{table}
\renewcommand{\arraystretch}{1.6}
\begin{center}
\begin{tabular}[h]{|l|l|l|}\hline
$p$ & the primitive distribution &\pageref{eqn.p}\\*[-5pt] & (a probability measure on $S =\{\I,\II\} \times [0,\infty) \times \N$) &\\
$\iota,\gamma,\xi$ & the three coordinate functions on $S$ &\pageref{def.rvs}\\*[-5pt] & (the active player, the capacity, the number of children) &\\
$q_{i}$ & the probability $p(\iota = i)$ & \pageref{def.activ}\\*[-5pt] & (player $i$'s activation probability) &\\
$p_{i}(k,n)$ & the probability $p(\{\iota = i\} \cap \{\gamma \geq k\} \cap \{\xi = n\})$ &\pageref{eqn.p}\\
$G_{i}(k,x)$ & generating function for the sequence $\{p_{i}(k,n)\}_{n \in \N}$ &\eqref{eqn.G}\\
$\omega$ & a game, the triple $(T_\omega,\iota_{\omega},\gamma_{\omega}$) &\pageref{def.game}\\
$\Omega$, $\mathbb{P}=\mathbb{P}_p$ & the space of games and a probability measure on $\Omega$& \pageref{def.space_of_games}, \eqref{eqn.measure1}, \eqref{eqn.measure2}\\
$E(i,k,n)$ & the event $\iota(\oslash) = i$, $\gamma(\oslash) \geq k$, $\xi(\oslash) = n$&\eqref{eqn.E(i,k,n)}\\
$\sigma_{i}$ & a strategy for player $i\in\{\I,\II\}$&\pageref{def.strategy}\\
$v=v_\omega$ & the value of the game $\omega$&\pageref{def.value}\\
$q_c$ & critical activation probability &\pageref{def.q_c.a}, \pageref{def.q_c.b}\\
$f$ & the value generating function (vgf)&\eqref{eqn.vgf}\\
$d(k)$ & a key parameter:&\eqref{eqn.d(k)}\\*[-5pt]& $\mathsf{E}_{p}(1_{\{\iota = \I\} \cap \{\gamma \geq k\}}\xi) + p(\{\iota = \II\} \cap \{\gamma \geq k\} \cap \{\xi = 1\})$&\\
$\omega_t$& the truncated game&\pageref{def.omega_t}\\
$v_t = v_{\omega_t}$& the value of the truncated game&\pageref{def.omega_t}\\
$\alpha(k)$& probability of the event $v < k$&\pageref{def.alpha}\\
$\beta(k)$& probability of the event $v \ge k$&\pageref{def.beta}\\
$\omega^*$& the ($k$-)conditional game&\pageref{def.omega*}\\
$p^*$& the primitive distribution&\pageref{def.p*}\\*[-5pt]&for the ($k$-)conditional game&\\
$f^*$& the vgf associated with $p^*$&\pageref{def.f*}\\
$\omega'$& the avoidance game&\pageref{def.omega'}\\\hline
\end{tabular}
\end{center}
\vspace{2mm}
\caption{Summary of notation\label{notation}}
\end{table}

\section{Random perfect information games}\label{secn.model}
Two players, I (Alice) and II (Bob), play against each other in a game that is chosen randomly by nature. The game is specified by a game tree, and by an assignment to its every node of the active player (either I or II), and of the capacity (a non-negative real number). The tree, the active players, and the capacities, are all generated by nature. Below we provide the technical details on the space of games, and of the measure from which a game is drawn. Informally, one could think of the number of children of a node, the active player, and the capacity being drawn by nature from a joint distribution independently across the nodes.

Upon observing the realization of the game, the two players decide on their strategies. The game is zero-sum, Player I being the maximizer. Player I's payoff is the smallest capacity encountered during play. A special case is the so-called escape model, where the capacity is set equal to zero whenever the node has no children. In the escape model, Player I's primary objective is to avoid childless nodes (i.e. ``escape to infinity"), since arriving at a childless node leads to the smallest possible payoff, zero. 

Each realization of the game is a zero-sum game with bounded upper semicontinuous payoffs. As is well known, each such game has a value. The distribution of the value is at the focus of the paper.\medskip 

\noindent\textbf{The primitive distribution:} Let $\N = \{0,1,\dots\}$ and $\N_+ = \{1,2,\dots\}$. 

The \textit{primitive distribution}, denoted by $p$\label{eqn.p}, is a joint probability distribution of the triple of random variables $(\iota, \gamma,\xi)$\label{def.rvs}, with $\iota$ taking values in the set $\{\I,\II\}$, $\gamma$ in $[0,\infty)$, and $\xi$ in $\N$. As will be clear shortly, the three random variables represent the active player, the capacity, and the number of children at a particular node of the game tree. Formally, we treat $p$ as a probability measure on the Borel subsets of $S = \{\I,\II\} \times [0,\infty) \times \N$\label{def.S} (where, of course, $\{\I,\II\}$ and $\N$ are equipped with the discrete topology), and $\iota$, $\gamma$, and $\xi$ as the three coordinate functions on $S$. For $(i,k,n) \in S$ let\label{def.activ}
\[q_{i} =: p(\{\iota = i\})\] and 
\begin{equation}\label{eqn.p}
p_{i}(k,n) =: p(\{\iota = i\} \cap \{\gamma \geq k\} \cap \{\xi = n\})\,.
\end{equation} 
The number $q_i$ is called \textit{player $i$'s activation probability}. Occasionally, we also write simply $q$ for $q_\I$ (and $1 - q$ for $q_\II$). We call the marginal of $p$ on $\xi$ \textit{offspring distribution}.  

Define, for $x \in [0,1]$:  
\begin{equation}\label{eqn.G}
G_{i}(k,x) =: \mathsf{E}_{p}(1_{\{\iota = i\} \cap \{\gamma \geq k\}} x^{\xi}) = \sum_{n \in \N}p_{i}(k,n)x^n\,,
\end{equation}
where $\mathsf{E}_{p}$ is the expectation with respect to the measure $p$. The functions $G_\I$ and $G_\II$ completely characterize the probability measure $p$. Note that $G_{i}(k,x)$ is the generating function for the sequence $\{p_{i}(k,n)\}_{n \in \N}$.

At this point, we allow for any joint distribution of $\iota$, $\gamma$, and $\xi$. In the sequel, however, we will attend closely to two classes of models. 

The first special case of importance is the escape model. We say that $p$ is an \textit{escape model} if $p(\{\gamma > 0\} \cap \{\xi = 0\}) = 0$ and $p(\{\gamma < k\}) > 0$ for each $k > 0$. The first condition imposes that all childless nodes of the game tree have zero capacity. By virtue of this condition, Player I's primary objective in the escape model is making sure that the game never stops, for reaching an end node results in the lowest possible payoff, zero. The second condition (which could be equivalently stated as saying that the essential infimum of $\gamma$ is $0$) is, in a sense, without loss of generality. If $p(\{\xi = 0\}) > 0$, it is implied by the first condition. And if $p(\{\xi = 0\}) = 0$, one could redefine the capacity by subtracting its essential infimum, obtaining a strategically equivalent specification.

The second important class of models are activation-independent models. We say that $p$ is \textit{activation-independent} if the random variable $(\gamma,\xi)$ is independent of $\iota$. Note that an activation-independent model still permits any joint distribution of $\gamma$ and $\xi$. Activation-independent models lend themselves to the study of comparative statics (of the distribution of the value) with respect to the activation probability of player I. To study comparative statics we fix a marginal of $p$ on $(\gamma,\xi)$ and think of $p$ as being parameterized by its marginal on $\iota$, that is, by $q$.\label{def.q}\medskip  

\noindent\textbf{Trees:} Let $H  = \cup_{n \in \N}\N_+^n$ denote the set of finite sequences of positive natural numbers, including the empty sequence $\oslash$. The length of a sequence $h \in H$ is the number $n \in \N$ such that $h \in \N_+^n$; in particular $\oslash$ has length $0$. For a sequence $h = (j_0,\dots, j_n)$ of length $n+1$, the sequences $\oslash$, $(j_0)$, $(j_0,j_1), \ldots, (j_0,\dots, j_n)$ are said to be the prefixes of $h$.  

A tree is a subset $T$ of $H$ containing $\oslash$ such that whenever a non-empty sequence is an element of $T$, all its prefixes are elements of $T$ as well. An infinite branch of $T$ is an infinite sequence $(j_0, j_1, \ldots) \in \N_{+}^{\N}$ such that $(j_0, \dots, j_t) \in T$ for every $t \in \N_+$. The boundary of the tree $T$, denoted $\partial T$, is the set of infinite branches of $T$.

A tree $T$ is an ordered tree if for each $h \in T$ there is a natural number $\xi(h) \in \N$ such that for $j \in \N_+$ the sequence $(h,j)$ is an element of $T$ if and only if $1 \leq j \leq \xi(h)$. Thus $\xi(h)$ is the number of children of the node $h$ in $T$; if $\xi(h) = 0$, then $h$ is an end node of $T$. Trees that are not ordered arise naturally in our setup as subsets of a given ordered tree. However, one can always obtain an ordered tree from a given tree by renaming its nodes in an order preserving manner. For a tree $T$ we let $o(T)$ denote the corresponding ordered tree, and $o : T \to o(T)$ be the corresponding bijection.\medskip 

%A metric on $\mathscr{T}$ is defined as follows: For $n \in \N$ let $T_n$ denote the set of sequences in $T$ of length of at most $n$. Define $d(T,T') =: \inf\{2^{-n}: n \in \N\text{ such that }T_n = T_n'\}$. The set $\mathscr{T}$ with the metric $d$ is a Polish space.
%One introduces a topology on the set of all labelled trees with the subbase consisting of the sets of the form $\{T: h \in T\}$, where $h \in H$.

\noindent\textbf{Games:}\label{def.space_of_games} We formally define a game as a particular type of a marked ordered tree, where the markings on a tree represent the active player and the capacity. The formalism of marked trees (``arbre marqué'') is borrowed from Neveu \cite{Neveu86}. 

A \textit{game} $\omega$ is a triple $(T_{\omega}, \iota_{\omega}, \gamma_{\omega})$ where $T_{\omega}$ is an ordered tree, and $\iota_{\omega}$ and $\gamma_{\omega}$ are functions on $T_{\omega}$ with values in $\{\I,\II\}$ and in $[0,\infty)$, respectively. The three elements of the game are, respectively, the \textit{game tree}, the assignment of the \textit{active player} to the nodes of the game tree, and the assignment of the \textit{capacity} to the nodes of the game tree. For $h \in T_{\omega}$ we let $\xi_{\omega}(h)$ denote the number of children of the node $h$ in the tree $T_{\omega}$. Let $\Omega$ be the set of games.\label{def.game} Given a game $\omega \in \Omega$ and a node $h \in T_{\omega}$, one defines a subgame $\omega(h)$ of $\omega$ starting at the node $h$. Formally $\omega(h) =: (s^{-1}[T_{\omega}], \iota_{\omega} \circ s, \gamma_{\omega} \circ s)$ where $s : H \to H$ is a shift operator given by $s(h') = (h,h')$. 

We endow $\Omega$ with a topology generated by the subbase consisting of sets of the form $\{\omega \in \Omega: h \notin T_{\omega}\}$ and $\{\omega \in \Omega: h \in T_{\omega},\,\iota_{\omega}(h) = i,\,k_0 < \gamma_{\omega}(h) < k_1\}$ where $h \in H$, $i \in \{\I,\II\}$, and $k_0$ and $k_1$ are rational numbers. The space $\Omega$ is Polish.

Note that the map $(\iota(\oslash),\gamma(\oslash),\xi(\oslash)) : \Omega \to S$ given by $\omega \mapsto (\iota_{\omega}(\oslash), \gamma_{\omega}(\oslash), \xi_{\omega}(\oslash))$ is a continuous function. The map $\omega \to \omega(1)$ is a continuous map from $\{\omega \in \Omega: \xi_{\omega}(\oslash) \geq 1\}$, a clopen subset of $\Omega$, into $\Omega$. For $(i,k,n) \in S$ let 
\begin{equation}\label{eqn.E(i,k,n)}
E(i,k,n) = \{\omega \in \Omega: \iota_{\omega}(\oslash) = i, \gamma_{\omega}(\oslash) \geq k, \xi_{\omega}(\oslash) = n\}\,.
\end{equation}

As is usual, we drop the subscript $\omega$ from our notation whenever this does not lead to a confusion.\medskip

\noindent\textbf{A measure on the space of games:}\label{def.measure} The following result is essentially due to Neveu \cite{Neveu86}: There exists a unique measure $\mathbb{P} = \mathbb{P}_p$ on the Borel subsets of $\Omega$ satisfying the following equalities:
\begin{align}
&\mathbb{P}(E(i,k,0) ) = p_i(k,0),\label{eqn.measure1}\\
&\mathbb{P}(E(i,k,n) \cap \bigcap_{j = 1}^{n}\{\omega(j) \in B_{j}\}) = p_i(k,n) \prod_{j = 1}^{n}\mathbb{P}(B_{j}),\label{eqn.measure2}
\end{align}
where $i \in \{\I,\II\}$, $k \geq 0$, $n \geq 1$, and $B_1, \dots, B_n$ are Borel subsets of $\Omega$.

The measure with the desired property can be constructed from a countable array of independent copies of the random variable $(\iota,\gamma,\xi)$ on $S$. To do so consider the set $\Pi = S^{H}$ endowed with the product topology; we think of an element $\pi$ of $\Pi$ as a triple $(\xi_{\pi}, \iota_{\pi}, \gamma_{\pi})$ of functions, mapping $H$ into $\N$, $\{\I,\II\}$, and $[0,\infty)$, respectively. Let $\otimes_{H}p$ be the product measure on $\Pi$ with the marginal on each coordinate equal to $p$. This defines a measurable space.

Given $\pi \in \Pi$ define the tree $T_{\pi}$ recursively: the empty sequence $\oslash$ is an element of $T_{\pi}$. If a sequence $h$ is an element of $T_{\pi}$, then for every $j \in \N_+$ the sequence $(h,j)$ is an element of $T_{\pi}$ if and only if $1 \leq j \leq \xi_{\pi}(h)$. Now define a map $g : \Pi \to \Omega$ by letting $g(\pi) = (T_{\pi}, \iota_{\pi}|_{T_{\pi}}, \gamma_{\pi}|_{T_{\pi}})$. The map $g$ is Borel measurable, and its distribution satisfies \eqref{eqn.measure1}--\eqref{eqn.measure2}.

One can give an alternative definition of $\mathbb{P}$ as the unique Borel probability measure on $\Omega$ satisfying the following two conditions:
\begin{itemize}
\item[{\rm [1]}] The random variable $(\iota(\oslash),\gamma(\oslash),\xi(\oslash))$ is distributed according to $p$. 
\item[{\rm [2]}] For each $(i,k,n) \in S$ where $n \in \N_+$, if $p_i(k,n) > 0$, then the random variables $\omega(1), \dots, \omega(n)$ are independent under the conditional measure $\mathbb{P}(\cdot\mid E(i,k,n))$, and each is distributed according to $\mathbb{P}$.
\end{itemize}

The measure $\mathbb{P}$ on the space of games is a natural generalization of the Galton-Watson measure on the space of ordered trees. In fact, the marginal of $\mathbb{P}$ on $T_{\omega}$ is a Galton-Watson measure.\medskip
%\begin{align*}
%q_{i} &=: p(\{\iota = i\}),\\p_{i}(k,n) &=: p(\{\iota = i\} \cap \{\gamma \geq k\} \cap \{\xi = n\}).
%\end{align*}

\noindent\textbf{How a game is played:} Consider a game $\omega \in \Omega$. The play in $\omega$ starts at the root $\oslash$.  Suppose that at some stage of the game a node $h$ of $T_\omega$ has been reached. If $h$ is an end node, the game ends. Otherwise, the active player $\iota_\omega(h)$ chooses one of the children $j \in \{1, \dots, \xi_\omega(h)\}$ of $h$, and the node $(h,j)$ of $T_\omega$ is reached at the next stage. Thus, a play of the game either leads to an end node of $T_\omega$, or induces an infinite branch of $T_\omega$. Let $h_0 = \oslash, h_1,h_2,\dots$ be a sequence, finite or infinite, of nodes of $T_\omega$ successively visited in the course of the game. The payoff to Player I is defined as $\inf\{\gamma_\omega(h_0), \gamma_\omega(h_1), \ldots\}$. Player I's goal is to maximize the payoff, and Player II's goal is to minimize it.

One can view the payoff function in the game $\omega$ as defined on the Baire space $\N_+^\N$, the space of infinite sequences of positive natural numbers, equipped with the product topology. Under this interpretation, the payoff is a bounded upper semicontinuous function.\medskip 

\noindent\textbf{Strategies:} A strategy\label{def.strategy} for Player I in the game $\omega$ is a function $\sigma_\I$ that assigns a number $\sigma_\I(h) \in \{1, \ldots, \xi_{\omega}(h)\}$ to each node $h \in T_{\omega}$ with $i_{\omega}(h) = \I$ and $\xi_{\omega}(h) \geq 1$. The interpretation is that at node $h$, the strategy $\sigma_\I$ recommends Player I to move to the child $(h, \sigma_\I(h))$ of $h$. A strategy $\sigma_\II$ for Player 2 is defined in a similar way. A pair of strategies $(\sigma_\I, \sigma_\II)$ either leads to an end node of $T_{\omega}$ eventually, or it induces an infinite branch of $T_{\omega}$. Let $u_{\omega}(\sigma_\I,\sigma_\II)$ denote the corresponding payoff.\medskip

\noindent\textbf{The value and $k$-optimal strategies:} As a consequence of Martin's determinacy theorem (\cite{Martin75}), the game $\omega$ has a value\label{def.value}:
\[v_{\omega} =: \sup_{\sigma_\I}\inf_{\sigma_\II}u_{\omega}(\sigma_\I,\sigma_\II) = \inf_{\sigma_\II}\sup_{\sigma_\I}u_{\omega}(\sigma_\I,\sigma_\II).\]
The value is the highest payoff that player I can guarantee to receive; at the same time, it is the lowest payoff that Player II can force upon Player I.

Let $k \geq 0$. We say that Player I's strategy $\sigma_\I$ is $k$-\textit{optimal} in $\omega$ if $u_{\omega}(\sigma_\I, \sigma_\II) \geq k$ for each Player II's strategy $\sigma_\II$ in $\omega$. In each game $\omega \in \Omega$, Player I has a $v_{\omega}$-optimal strategy\footnote{Note a slight departure from the standard terminology: what we call a $v_{\omega}$-optimal strategy would usually be called an optimal strategy.}. Likewise, Player II's strategy $\sigma_\II$ is said to be $k$-\textit{optimal} in $\omega$ if $u_{\omega}(\sigma_\I,\sigma_\II) \leq k$ for each Player I's strategy $\sigma_\I$. Player II might not have a $v_{\omega}$-optimal strategy, but he does a $(v_{\omega} + \epsilon)$-\textit{optimal} strategy for each $\epsilon > 0$. If the capacity is supported on a finite set, then Player II also has a $v_{\omega}$-optimal strategy (see e.g. Laraki, Maitra, and Sudderth \cite{Laraki}).

For $h \in T_{\omega}$ we write $v_{\omega}(h)$ to denote $v_{\omega(h)}$, the value of the subgame $\omega(h)$ of $\omega$.\medskip

\section{Two examples}\label{secn.examples}
This section introduces two examples that will serve as the illustration throughout the paper. The data reported here is based on the main results derived in the following section (see Subsection \ref{subsecn.main}). We hope to convey the flavour of our main findings, and help the reader anticipate the developments in the rest of the paper.

Note that both examples are instances of an activation-independent escape model. 

\begin{exl}\label{exl.fraclin}\rm
Suppose that the number of children $\xi$ has the geometric distribution: $p(\xi = n) = (1-l)l^n$, where $0 < l < 1$. Recall that $\mathsf{E}_{p}(\xi) = l/(1 - l)$. 

To complete the description of the example, assume that $\iota$ and $\xi$ are independent, and that $\gamma = 1$ whenever $\xi > 0$ and $\gamma = 0$ if $\xi = 0$. This specification of the capacity is a particularly important special case of the model. Under this specification the sole goal of Player I is to avoid end nodes of the tree. The value of the game is either $0$ or $1$, and it is $0$ precisely when Player II can force a play to reach an end node. We say that Player I wins the game if $v=1$, and that Player II wins the game if $v=0$. We let $\alpha$ denote the probability of the event $\{v = 0\}$ and and by $\beta$ the probability of $\{v = 1\}$. 

Consider first the non-adversarial scenario, the scenario under which all nodes are assigned to player I (i.e. $q = 1$). Of course, in the non-adversarial case $v = 0$ precisely when the game tree $T_{\omega}$ has no infinite branches, that is, in the event of ``extinction'' of the game tree. Since under the measure $\PP$, the distribution of the game tree is a Galton-Watson measure, classical results on the branching processes apply (e.g. Athreya and Ney \cite{Athreya}): the probability of extinction of the game tree is $\alpha = 1$ if $l/(1-l) \leq 1$ and is $\alpha = (1-l)/l$ otherwise. 

Turning to the general case, Figure \ref{fig.fraclin} depicts $\alpha$ as a function of $q$ for $l = 0.6$ (blue) and $l = 0.9$ (red). Predictably, $\alpha$ is decreasing in $q$, with $q = 1$ corresponding to the non-adversarial scenario. The figure reveals an interesting phase transition: if Player I controls too few nodes, she has no chance of winning the game. The critical activation probability $q_c$ for player I is $0.6032$ for $l = 0.6$, and $0.1021$ for $l = 0.9$. \label{def.q_c.a}

We find that $\alpha < 1$ (or, equivalently, that $\beta > 0$) if and only if $q > q_{c}$, where the critical activation probability is given by\footnote{The general expression for the critical activation probability $q_c$ is \eqref{eqn.criticalq}. The expression for $\alpha$ follows by Theorem \ref{thm.fixed} using Example \ref{exl.vgf}.}: 
\[q_c = \frac{(1-l)(1-l+l^2)}{l^2(2 - l)}.\] 
Whenever $q > q_c$, the probability of $\{v = 0\}$ is given by  
\[\alpha = \tfrac{1}{2}(2-l)(1-q) + \tfrac{1}{2}\sqrt{4\frac{(1-l)^2}{l^2} + (2-l)^2(1-q)^2}.\]
%We have the following bounds: $\max\{a, b\} \leq \alpha \leq a + b$. 

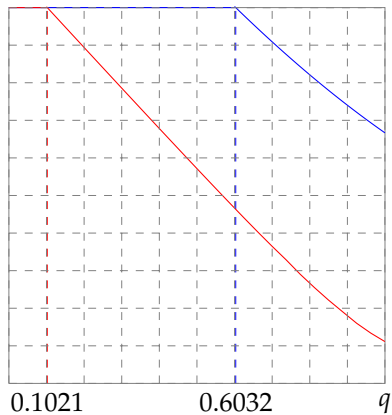
\begin{figure}
\caption{The probability $\alpha = \mathbb{P}(v = 0)$ in Example \ref{exl.fraclin} as a function of $q$ for $l = 0.6$ (blue) and $l = 0.9$ (red).}
\label{fig.fraclin}
\begin{center}
\begin{tikzpicture}[domain=0:1,scale=5]
\draw[very thin,color=gray](0,0)--(1,0)--(1,1)--(0,1)--(0,0);
\draw[color=blue](0,1)--(0.6,1);
\node[below] at(1,0){$q$};
\draw[color=blue, domain=0.6:1] plot (\x,{0.5*(2-0.6)*(1-\x) + 0.5*(4*(0.4/0.6)^2 + ((2-0.6)*(1-\x))^2)^(0.5)});
\draw[step=0.2, color=blue, line width=0.2pt, dashed](0.6032,0)--(0.6032,1);
\node[below] at(0.6032,0){$0.6032$};
\draw[color=red](0,1)--(0.1,1);
\draw[color=red, domain=0.1:1] plot (\x,{0.5*(2-0.9)*(1-\x) + 0.5*(4*(0.1/0.9)^2 + ((2-0.9)*(1-\x))^2)^(0.5)});
\draw[step=0.2, color=red, line width=0.2pt, dashed](0.1021,0)--(0.1021,1);
\node[below] at(0.1021,0){$0.1021$};
\draw [step=0.1, gray, line width=0.2pt, dashed](0,0)grid(1,1);
\end{tikzpicture}
\end{center}
\end{figure}
\end{exl}

\begin{exl}\label{exl.tree}\rm
Let $n \geq 2$ and suppose that under the measure $p$, $\xi = n$ almost surely (so that the game tree $T_{\omega}$ is a complete $n$-ary tree), that $\gamma$ and $\iota$ are independent, and that $\gamma$ is uniformly distributed on $[0,1]$. Clearly, the value $v_{\omega}$ of any game $\omega$ is an element of $[0,1]$. We are particularly interested in the essential supremum of the value.

First consider the non-adversarial scenario ($q = 1$). In the non-adversarial scenario the probability that $\{v_{\omega} \geq k\}$ is positive if and only if $(1-k)n > 1$. This implies that the essential supremum of the value is $1 - 1/n$. We briefly describe the rationale behind this conclusion. If Player I controls all nodes, then $v_{\omega} \geq k$ if and only if the tree $T_{\omega}$ has an infinite branch that passes only through the nodes with a capacity no smaller than $k$. Thus we may remove all the children of any node with a capacity smaller than $k$, and let $T_{\omega}^{k}$ be an infinite component of the root of the remaining subgraph. Then $\{v_{\omega} \geq k\}$ is exactly the event of non-extinction of the tree $T_{\omega}^{k}$. One can see that the tree $T_{\omega}^{k}$ is distributed according to a Galton-Watson measure generated by an offspring distribution  with mean $(1-k)n$. We may thus conclude that $T_{\omega}^{k}$ has an infinite branch with positive probability if and only if $(1-k)n > 1$. 

Turning to the general case, we find that, for $k > 0$, the probability that $\{v \geq k\}$ is positive if and only if $(1-k)nq > 1$. In particular, if $q \leq 1/n$, then $v = 0$ almost surely, and if $1/n < q$, then the essential supremum of the value is $1 - 1/nq$. Interestingly, if player I controls too few nodes, then player II is able to force the play to go through the nodes with a vanishingly small capacity, thus ensuring the payoff of zero. 

Let us define the $k$--critical activation probability for Player I to be $q_c(k) = 1/(1-k)n$ if $(1-k)n > 1$ and to be $1$ otherwise. Then $\mathbb{P}(v \geq k) > 0$ if and only if $q > q_c(k)$. For the binary and the ternary trees, the probability of the event $\{v < k\}$ allows for a simple closed-form solution.\footnote{These expressions can be derived easily using Example \ref{exl.vgf} and Theorem \ref{thm.fixed}.}  For the binary tree we have:

\begin{equation}\label{eqn.binary}
\mathbb{P}(v < k) = \begin{cases}
1 &\text{if }2(1-k)q \leq 1\\
\displaystyle\frac{k}{(1-k)}\frac{1}{(2q - 1)}&\text{if }1 < 2(1-k)q.
\end{cases}
%\quad
%\mathbb{P}(v \geq k) = \begin{cases}
%0 &\text{if }2pq \leq 1\\
%\tfrac{2pq-1}{2pq - p}&\text{if }2pq > 1
%\end{cases}
\end{equation}
For the ternary tree:
\begin{equation}\label{eqn.ternary}
\mathbb{P}(v < k) = \begin{cases}
1 &\text{if }3(1-k)q \leq 1\\
\displaystyle\tfrac{1}{2}(2 - 3q) +\tfrac{1}{2}\sqrt{(2- 3q)^2 + 4\frac{k}{(1-k)}}&\text{if }1 < 3(1-k)q.
\end{cases}
\end{equation}

The left panel of Figure \ref{fig.tree} displays $\mathbb{P}(v < 0.05)$ for the binary tree (in blue) and for the ternary tree (in red) as a function of $q$. Note in particular the $0.05$-critical probability: $q_c(0.05) = 0.5263$ in the case of binary tree, and $q_c(0.05) = 0.3509$ for the ternary. It emerges from the figure that enlarging the tree does not necessarily benefit Player I: if $q = 0.7$, then the probability that $v < 0.05$ goes up from $0.1316$ for the binary tree to $0.1848$ for the ternary. 

The right panel of Figure \ref{fig.tree} depicts $\mathbb{P}(v < k)$ as a function of $k$ (that is, the cumulative distribution function of the value) for the binary tree (blue) and the ternary tree (red), assuming that $q = 0.7$. Note the essential supremum of the value: $0.2857$ for the binary tree, and $0.5238$ for the ternary.
  
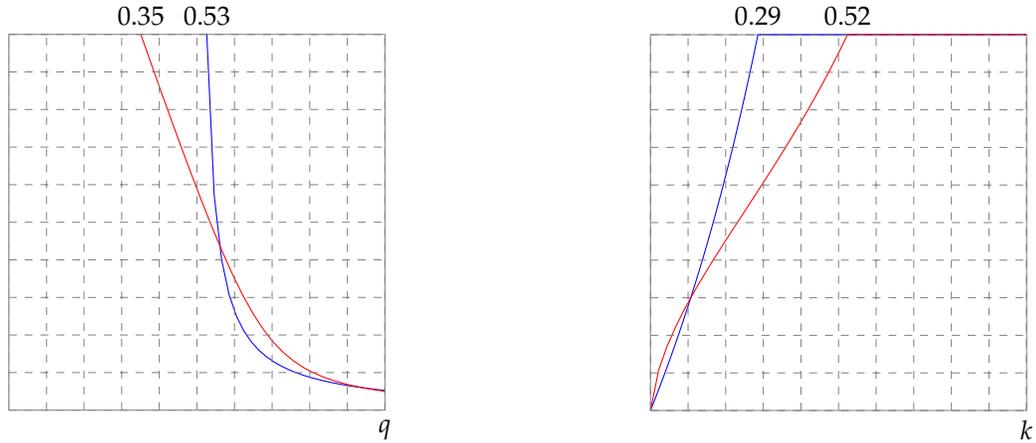
\begin{figure}[h!]
\caption{The probability $\mathbb{P}(v < k)$ in Example \ref{exl.tree}.}
\label{fig.tree}
\begin{subfigure}{0.5\textwidth}
\begin{center}
\begin{tikzpicture}[domain=0:1,scale=5]
\draw[very thin,color=gray](0,0)--(1,0)--(1,1)--(0,1)--(0,0);
\draw[color=blue, domain=0.5263:1] plot (\x,{(1/19)*(2*\x - 1)^(-1)});
%\draw[step=0.2, color=blue, line width=0.2pt, dashed](0.5263,0)--(0.5263,1);
\node[above] at(0.5263,1){0.53};
\draw[color=red, domain=0.3508:1] plot (\x,{0.5*(2 - 3*\x) + 0.5*((2 - 3*\x)^2 + 4*(1/19))^(0.5)});
%\draw[step=0.2, color=red, line width=0.2pt, dashed](0.3509,0)--(0.3509,1);
\node[above] at(0.3509,1){0.35};
\node[below] at(1,0){$q$};
\draw [step=0.1, gray, line width=0.2pt, dashed](0,0)grid(1,1);
\end{tikzpicture}
\end{center}
\subcaption{The probability of $\{v < 0.05\}$ for the binary (blue) and ternary (red) trees as a function of $q$.}
\end{subfigure}\hspace{1cm}
\begin{subfigure}{0.5\textwidth}
\begin{center}
\begin{tikzpicture}[domain=0:1,scale=5]
\draw[very thin,color=gray](0,0)--(1,0)--(1,1)--(0,1)--(0,0);
\draw[color=blue, domain=0:0.2857] plot (\x,{\x/((1 - \x)*(2*0.7 - 1))});
\draw[color=blue] (0.2857,1)--(1,1);
\node[above] at(0.2857,1){0.29};
\node[below] at(1,0){$k$};
\draw[color=red, domain=0:0.5238] plot (\x,{0.5*(2 - 3*0.7) + 0.5*((2 - 3*0.7)^2 + 4*\x/(1-\x))^(0.5)});
\draw[color=red] (0.5238,1)--(1,1);
\node[above] at(0.5238,1){0.52};
\draw[step=0.1, gray, line width=0.2pt, dashed](0,0)grid(1,1);
\end{tikzpicture}
\end{center}
\subcaption{The cumulative distribution of $v$ for the binary (blue) and the ternary (red) trees, $q = 0.7$.}
\end{subfigure}
\end{figure}
\end{exl}

\section{The distribution of the value}\label{secn.distribution}
In this section we derive the main result of the paper: a fixed point characterization of the distribution of the value. In the first two subsections  we introduce our tools: a family of the so-called value generating functions, and the truncated games. The last subsection states the main results.

The value generating functions reflect the recursive nature of the value. Effectively, they represent the Shapley operator. Intuitively, they map the distribution of the value in the next period to that in the current period. The $t$th iterate of the value generating functions at $0$ determine the distribution of the value in the $t$-truncated game. This, and the fact that the value of the truncated games converge (from above) to the value of the infinite game eventually yield the fixed point characterization of the distribution of the value.   

\subsection{The value generating function}
Define the \textit{value generating function} (vgf) $f : [0,\infty) \times [0,1] \to [0,1]$ associated with the primitive distribution $p$ by 
\begin{equation}\label{eqn.vgf}
f(k,x) = 1 - G_\I(k,0) - G_\I(k,1) + G_\I(k,x) - G_\II(k,1-x).
\end{equation}
%\[f(k,x) = 1 - p_\I(k,0) - \sum_{n \in \N}p_\I(k,n) + \sum_{n \in \N}p_\I(k,n)x^{n} - \sum_{n \in \N}p_\II(k,n)(1-x)^{n}\,,
%\end{equation}

We write $f_k : [0,1] \to [0,1]$ to denote the function $x \mapsto f(k,x)$.

\begin{exl}\label{exl.vgf}\rm
In the setup of Example \ref{exl.fraclin} we have
\begin{equation}\label{eqn.fraclinu}
G_{i}(1,x) = q_{i}\frac{(1-l)lx}{1-lx},
\end{equation}
from which we derive the vgf:
\[f_{1}(x) = 1 - l(1-x)\Big(q_\I\frac{1}{1 - l x} + q_\II\frac{1 - l}{1 - l + l x}\Big).\]
%\[f(x) = 1 - q_\Il + q_\I(1 - l)\frac{l x}{1 - l x} - q_\II(1 - l)\frac{l - l x}{1 - l + l x}.\]

The function is displayed in the left panel of Figure \ref{fig.vgf} for $l = 0.9$ and $q = 0.5$ (blue), for $l = 0.6$ and $q = 0.5$ (red). Especially the first of these (the blue) displays a feature characteristic of all vgfs: it is concave on an interval $[0,c]$ for some $0< c < 1$ and convex on $[c,1]$. 

In the setup of Example \ref{exl.tree} the vgf is given by
\begin{equation}\label{eqn.treesu}
G_{i}(k,x) = (1-k)q_{i}x^n,
\end{equation}
and hence
\[f_{k}(x) = 1 - (1-k)q_\I + (1-k)q_\I x^n - (1-k)q_\II(1-x)^n.\]
It is pictured in the right panel of Figure \ref{fig.vgf} with $k = 0.05$ and $q = 0.7$ for $n = 3$ (blue), $n = 4$ (orange), and $n = 8$ (red). 

\begin{figure}[h]
\caption{vgf}
\label{fig.vgf}
\begin{subfigure}{.5\textwidth}
\begin{center}
\begin{tikzpicture}[domain=0:1,scale=5]
\draw[very thin,color=gray](0,0)--(1,0)--(1,1)--(0,1)--(0,0);
\draw[color=blue, domain=0:1] plot (\x,{1 - 0.9*(1-\x)*(0.5/(1 - 0.9*\x) + 0.5*(1-0.9)/(1 - 0.9 + 0.9*\x)});
\draw[color=red, domain=0:1] plot (\x,{1 - 0.6*(1-\x)*(0.5/(1 - 0.6*\x) + 0.1*0.5/(1 - 0.6 + 0.6*\x)});
\node[below] at(1,0){$x$};
\draw [step=0.1, gray, line width=0.2pt, dashed](0,0)--(1,1);
\end{tikzpicture}
\caption{The vgf in Example \ref{exl.fraclin} for $l = 0.9$ and $q = 0.5$ (blue), and for $l = 0.6$ and $q = 0.5$ (red).}
\end{center}
\end{subfigure}
\hspace{1cm}
\begin{subfigure}{.5\textwidth}
\begin{center}
\begin{tikzpicture}[domain=0:1,scale=5]
\draw [step=0.1, gray, line width=0.2pt, dashed](0,0)--(1,1);
\draw[very thin,color=gray](0,0)--(1,0)--(1,1)--(0,1)--(0,0);
\draw[color=blue, domain=0:1] plot (\x,{1 - (1-0.05)*0.7 + (1-0.05)*0.7*\x^3 - (1-0.05)*(1-0.7)*(1-\x)^3)});
\draw[color=orange, domain=0:1] plot (\x,{1 - (1-0.05)*0.7 + (1-0.05)*0.7*\x^4 - (1-0.05)*(1-0.7)*(1-\x)^4)});
\draw[color=red, domain=0:1] plot (\x,{1 - (1-0.05)*0.7 + (1-0.05)*0.7*\x^8 - (1-0.05)*(1-0.7)*(1-\x)^8)});
\node[below] at(1,0){$x$};
\end{tikzpicture}
\caption{The vgf in Example \ref{exl.tree} with $k = 0.05$ and $q = 0.7$ for $n = 3$ (blue), $n = 4$ (orange), and $n = 8$ (red).}
\end{center}
\end{subfigure}
\end{figure}
\end{exl}

Define 
\begin{equation}\label{eqn.d(k)}
d(k) = \mathsf{E}_{p}(1_{\{\iota = \I\} \cap \{\gamma \geq k\}}\xi) + p(\{\iota = \II\} \cap \{\gamma \geq k\} \cap \{\xi = 1\}).
\end{equation}
This quantity turns out to be one of the key parameters of our model. 

The following lemma summarizes the relevant properties of the vgf.

%\begin{equation}\label{eqn.d} 
%d(k) = \mathsf{E}_{p}(1_{\{\iota = \I\} \cap \{\gamma \geq k\}}\xi) + p(\{\iota = \II\} \cap \{\gamma \geq k\} \cap \{\xi = 1\}). 
%\end{equation}

\begin{lemma}\label{thm.vgf} 
Let $k \geq 0$.
\begin{itemize}
\item[(i)] The function $f_k$ is continuous and non-decreasing on $[0,1]$. It is differentiable any number of times at any point of $(0,1)$. 
\item[(ii)] It holds that 
\begin{align*}
f_{k}(0) &= p(\{\gamma < k\}),\\
f_{k}(1) &= p(\{\gamma < k\} \cup \{\xi \geq 1\}),\\ 
\lim_{x \uparrow 1}\tfrac{\partial f_{k}}{\partial x}(x) &= d(k).
\end{align*}
\item[(iii)] There exists a point $c \in [0,1]$ such that $f_{k}$ is concave on $[0,c]$ and convex on $[c,1]$.  
\item[(iv)] Suppose that $f_k$ is not the identity map. If $s \in (0,1)$ is a fixed point of $f_k$, then $x < f_k(x)$ for each $x \in (0,s)$ and $f_k(x) < x$ for each $x \in (s,1)$. In particular, $f_k$ has at most one fixed point in $(0,1)$.   
\item[(v)] The function $f_k$ has no fixed point in $[0,1)$ if and only if all of the following three conditions are satisfied: $f_k(0) > 0$, $f_k(1) = 1$, and $d(k) \leq 1$.
\end{itemize}
\end{lemma}

\begin{proof}
We write $f=f_k$, and $G_{i}(x)$ for $G_{i}(k,x)$.

\textsc{Claim (i)}: The function $G_{i}$ is a generating function for the sequence $\{p_{i}(k,n)\}_{n \in \N}$. It follows that (Grimmett and Stirzaker \cite[\S5.1]{Grimmett}) both these functions are continuous on $[0,1]$, and they may be differentiated term by term any number of times at any point $x \in (0,1)$. It holds that $f'(x) = G_\I'(x) + G_\II'(1-x) \geq 0$ for each $x \in (0,1)$.

\textsc{Claim (ii)}: By a direct computation.

\textsc{Claim (iii)}: First we argue that either $f$ is linear, or it has at most one inflection point in $(0,1)$, that is, only one point $x \in (0,1)$ such that $f''(x) = 0$. 

For each $x \in (0,1)$ it holds that $f'''(x) = G_\I'''(x) + G_\II'''(1-x)$, where $G_{i}'''(x) \geq 0$ for both $i \in \{\I,\II\}$. In particular, $f''$ is non-decreasing on $(0,1)$. If $f$ has two distinct inflection points in $(0,1)$, then $f''$ and hence also $f'''$ vanish on a non-degenerate interval, say $(c_0,c_1)$. Hence $G_\I'''(x) = 0$ and $G_\II'''(1-x) = 0$ for each $x \in (c_0,c_1)$. It follows that  $p(\{\gamma \geq k\} \cap \{\xi \geq 3\}) = 0$. But in this  case $f$ is either linear or quadratic. In the latter case it has no inflection points, proving the assertion.

If $f$ is a linear function, we can set $c = 0$. Now suppose that $f$ has at most one inflection point in $(0,1)$. Define $c$ to be $0$ if $f''(x) > 0$ for all $x \in (0,1)$, to be $1$ if $f''(x) < 0$ for all $x \in (0,1)$, and otherwise to be the unique point of $(0,1)$ such that $f''(c) = 0$. Recalling that $f''$ is non-decreasing, we conclude that $f$ is concave on $[0,c]$, and is convex on $[c,1]$.

\textsc{Claim (iv)}: The claim is trivial if $f$ is a linear function. So suppose that $f$ is not linear. By the earlier conclusion, $f$ has at most one inflection point in $(0,1)$. We distinguish two cases.

Case 1: $c \leq s$. We know that $f(s) = s$ and $f(1) \leq 1$, and that $f$ is convex on $[s,1]$. This implies that $f(x) \leq x$ for each $x \in [s,1]$. Moreover, if there existed a point $x \in (s,1)$ such that $f(x) = x$, the function $f$ would be the identity map on $[s,1]$, implying a continuum of inflection points, a contradiction.

We also have $x < f(x)$ for each $x \in [c,s)$. For otherwise convexity of $f$ on $[c,1]$ would imply that $f$ is the identity map on $[s,1]$, contradicting the fact that it has at most one inflection point in $(0,1)$. We also know that $0 \leq f(0)$ and that $c < f(c)$. Concavity of $f$ on $[0,c]$, now implies that $x < f(x)$ for each $x \in (0,c]$.

Case 2: $s \leq c$. One applies a similar reasoning to the intervals $[0,s]$, $[s,c]$, and $[c,1]$. 

\textsc{Claim (v), necessity}: Suppose that $f$ has no fixed point in $[0,1)$. Obviously then $0 < f(0)$. If there existed a point $x \in [0,1]$ such that $f(x) < x$, the intermediate value theorem would imply that $f$ has a fixed point in $(f(0),f(x))$, contradicting the supposition. Thus $x \leq f(x)$ for each $x \in [0,1]$. In particular $f(1) = 1$. If $c = 1$, then $f$ is concave on $[0,1]$. Then $xf'(x) \leq f(x) - f(0)$ for each $x \in (0,1)$. Letting $x \uparrow 1$ we obtain $d(k) \leq 1$. If $c < 1$, then $f$ is convex on $[c,1]$. In this case $f'(x)(1-x) \leq f(1) - f(x) \leq 1 - x$ for each $x \in (c,1)$, and thus $d(k) \leq 1$.

\textsc{Claim (v), sufficiency}: Suppose that $0 < f(0)$, $f(1) = 1$, and $d(k) \leq 1$. Towards a contradiction, let $s \in (0,1)$ be a fixed point of $f$. Clearly, this means that $f$ cannot be concave on $[0,1]$, hence $c < 1$, and so $f$ is convex on $[c,1]$. On the other hand, by Claim (iv), $f(x) < x$ for each $x \in (s,1)$. But then there is a point $x_0 \in (c,1)$ such that $1 < f'(x_0)$, and since $f'(x_0) \leq f'(x)$ for all $x \in (x_0,1)$, we obtain $1 < f'(x_0) \leq d(k)$, a contradiction.
\end{proof}

\subsection{The truncated game}
Given a game $\omega$ and time $t \in \N$, we define a version of the game $\omega$ played on a game tree truncated at time $t$. The idea is straightforward: the game $\omega_{t}$ lasts for no more than $t$ periods, and Player I's payoff is the smallest capacity along the nodes visited prior to the deadline $t$. Given a game $\omega = (T_{\omega}, \iota_{\omega}, \gamma_{\omega})$ define the \textit{truncated game} $\omega_{t}$\label{def.omega_t} to be the triple $(T_{\omega_{t}}, \iota_{\omega_{t}}, \gamma_{\omega_{t}})$, where $T_{\omega_{t}}$ is the subset of nodes of $T_{\omega}$ having the length of at most $t$, and $\iota_{\omega_{t}}$ and $\gamma_{\omega_{t}}$ are the restrictions of $\iota_{\omega}$ and $\gamma_{\omega}$, respectively, to $T_{\omega_{t}}$. In particular, $\omega_0$ has a tree consisting of a single node, namely the empty sequence $\oslash$.    

We first establish measurability of the value of a game $\omega$. We remark that the measurability of the value cannot be taken for granted. It is known that the value of a Borel-parameterized infinite perfect information game need not be Borel-measurable (Moschovakis \cite{Moschovakis}); counterexamples to measurability of the value have also be given in other contexts, for instance for simultaneous move games (Prikry and Sudderth \cite{PrikrySudderth}). Positive result below can be linked to the fact that Player I's payoff function is, in each game $\omega$, upper semicontinuous.

\begin{lemma}\label{thm.finite:meas}
Let $t \in \N_+$
\begin{itemize}
\item[(i)] The map $\Omega \to \Omega$ given by $\omega \mapsto \omega_t$ is continuous.
\item[(ii)] The map $v_t : \Omega \to \mathbb{R}$ defined by $\omega \mapsto v_{\omega_t}$, is continuous.
\item[(iii)] For each $\omega \in \Omega$, $v_{\omega_0} \geq v_{\omega_1} \geq \cdots$ is a non--increasing sequence converging to $v_{\omega}$. Consequently, the map $v : \Omega \to \mathbb{R}$, $\omega \mapsto v_{\omega}$, is upper semicontinuous, hence measurable.
\end{itemize}
\end{lemma}
\begin{proof}
Items (i) can be checked directly using the definition of the subbase for of the topology on $\Omega$. Item (ii) can be shown by an induction on $t$. We turn to item (iii).

Let $v_{\omega_{t}} = k$, and let $\sigma_\II$ be Player II's $k$-optimal strategy in the game $\omega_{t}$. Note that, $\omega_{t}$ being essentially a finite game, both players have an optimal strategy. The strategy $\sigma_\II$ guarantees that, within $t$ periods of time, the play visits a node with a capacity of at most $k$. Clearly then, the same strategy is $k$-optimal in $\omega_{t+1}$. Thus $v_{\omega_{t}} \geq v_{\omega_{t+1}}$.

Now let $v_{\omega} = k$. Also let $\epsilon>0$.

Firstly, we have $v_{\omega_{t}} \geq k$ for each $t \in \N$, since each $k$-optimal strategy of Player I in $\omega$ is also a $k$-optimal strategy in $\omega_{t}$.

To see that the sequence $v_{\omega_{t}}$ converges to $v_{\omega} = k$, fix some ($k+\epsilon$)-optimal strategy for Player II in $\omega$, say $\sigma_\II$. Let $W \subseteq T$ denote a tree consisting of the nodes $h \in T_{\omega}$ that (i) could be reached when Player II is using his strategy $\sigma_\II$, and (ii) have the property that $\gamma(h') > k + \epsilon$ for each prefix $h'$ of $h$. Then the tree $W$ is well--founded, i.e. it has no infinite branches. Since it is a locally finite tree, it is actually finite. Let $t$ be the height of $W$. We conclude that any play of the game $\omega$ consistent with $\sigma_\II$ reaches a node with a capacity of $k + \epsilon$ or less by period $t$. But this means that $\sigma_\II$ is a ($k+\epsilon$)-optimal strategy in $\omega_{t}$. Thus $v_{\omega_{t}} \leq k + \epsilon$, and hence $\inf_{t \in \N}v_{\omega_{t}} \leq k + \epsilon$. Since $\epsilon > 0$ is arbitrary, we have shown that $\inf_{t \in \N}v_{\omega_{t}} \leq k$.
\end{proof}

We turn to the probabilistic properties of the value. Recall that $\omega(h)$ denotes the subgame of the game $\omega$ starting at a node $h \in T_{\omega}$. Recall also the definition of the events $E(i,k,n)$ in \eqref{eqn.E(i,k,n)}.

\begin{lemma}\label{thm.finite:cond}
Let $t \in \N_{+}$.
\begin{itemize}
\item[(i)] It holds that $(\omega_t)(\ell) = (\omega(\ell))_{t-1}$ for $\omega \in \Omega$ and each $\ell \in \{1, \dots, \xi_{\omega}(\oslash)\}$.
\item[(ii)] Take $(i,k,n) \in S$ such that $p_i(k,n) > 0$. For $\ell \in \{1, \dots, n\}$ let $v_{t}(\ell) : E(i,k,n) \to \mathbb{R}$ denote the random variable $\omega \mapsto v_{(\omega_t)(\ell)}$. Under the conditional measure $\mathbb{P}(\cdot \mid E(i,k,n))$, the random variables $v_{t}(1),\dots,v_{t}(n)$ are independent, and each is distributed like the random variable $v_{t-1}$.
\end{itemize}
\end{lemma}
\begin{proof}
Item (i) follows immediately from the definition of the truncated game. Item (ii) follows from item (i), and the fact that under the conditional measure $\mathbb{P}(\cdot \mid E(i,k,n))$, the random variables $\omega(1),\dots,\omega(n)$ are independent, and each is distributed like the random variable $\omega$. 
\end{proof}

\begin{lemma}\label{thm.finite}
Take $k > 0$. Let $\alpha_{t} = \PP(v_{t} < k)$ and $\beta_{t} = \PP(v_{t} \geq k)$. Then the sequence $\alpha_{0} \leq \alpha_{1} \leq \cdots$ converges to $\PP(v < k)$. For each $t \in \N$ we have $\alpha_{t} = f_k^{t+1}(0)$, where $f_k^{t+1}$ denotes the $(t+1)$-fold iterate of $f_k$.
%\item[(ii)] For each $t \in \N$, it holds that
%\[\begin{aligned}
%&\mathbb{P}(\{v_{t} \geq k\} \cap \{\iota(\oslash) = \I\}) =& G_\I(k,1) - G_\I(k,\alpha_{t-1})\\
%&\mathbb{P}(\{v_{t} \geq k\} \cap \{\iota(\oslash) = \II\}) =& G_\II(k,\beta_{t-1}).
%\end{aligned}\]
\end{lemma}
\begin{proof}
By Lemma \ref{thm.finite:meas}, $\{v_{0} < k\} \subseteq \{v_{1} < k\} \subseteq \cdots$ is a non-decreasing sequence of events converging to $\{v < k\}$. The first claim follows.

Note that $v_t = v_{\omega_t} \leq \gamma_{\omega_t}(\oslash) = \gamma_{\omega}(\oslash) < k$ everywhere on the complement of the event $\cup\{E(i,k,n):i \in \{\I,\II\},n \in \N\}$. 

Consider the event $E(\I,k,0)$. Since $v_{t} = \gamma_{\omega}(\oslash) \geq k$ holds everywhere on $E(\I,k,0)$, we have
\[\mathbb{P}(v_{t} \geq k \mid E(\I,k,0)) = 1;\]
whenever $E(\I,k,0)$ has a positive probability.

Take $n \geq 1$ and consider the event $E(\I,k,n)$. We have
\[E(\I,k,n) \cap \{v_{t} < k\} = E(\I,k,n) \cap_{j = 1}^{n} \{v_{t}(j) < k\}.\]
Hence, using Lemma \ref{thm.finite:cond}, we compute:
\[\mathbb{P}(v_{t} < k \mid E(\I,k,n)) = \mathbb{P}(\cap_{j = 1}^{n} \{v_{t}(j) < k\} \mid E(\I,k,n)) = \prod_{\ell = 1}^{n} \mathbb{P}(v_{t-1} < k) = \alpha_{t-1}^{n},\]
whenever $E(\I,k,n)$ has a positive probability. Thus 
\begin{align*}
\mathbb{P}(\{v_{t} \geq k\} \cap \{\iota_{\omega}(\oslash) = \I\}) = &\sum_{n \in \N}\mathbb{P}(\{v_{t} \geq k\} \cap E(\I,k,n))\\
=&\sum_{n \in \N} \mathbb{P}(E(\I,k,n)) \mathbb{P}(\{v_{t} \geq k\} \mid E(\I,k,n))\\
=&p_\I(k,0) + \sum_{n \in \N}p_\I(k,n)(1 - \alpha_{t-1}^{n})\\
=&G_\I(k,0) + G_\I(k,1) - G_\I(k,\alpha_{t-1}).
\end{align*}

Consider the event $E(\II,k,0)$. Since $v_{t} = \gamma_{\omega}(\oslash) \geq k$ holds everywhere on $E(\II,k,0)$, we have
\[\mathbb{P}(v_{t} \geq k \mid E(\II,k,0)) = 1;\]
whenever $E(\II,k,0)$ has a positive probability. 

Let $n \geq 1$, and consider the event $E(\II,k,n)$. We have 
\[E(\II,k,n) \cap \{v_{t} \geq k\} = E(\II,k,n) \cap_{j = 1}^{n} \{v_{t}(j) \geq k\}.\]
Hence using Lemma \ref{thm.finite:cond} we obtain
\[\mathbb{P}(v_{t} \geq k \mid E(\II,k,n)) = \mathbb{P}(\cap_{j = 1}^{n} \{v_{t}(j) \geq k\} \mid E(\II,k,n)) = \prod_{j=1}^{n} \mathbb{P}(v_{t-1} \geq k) = \beta_{t-1}^{n};\]
whenever $E(\II,k,n)$ has a positive probability. Thus
\begin{align*}
\mathbb{P}(\{v_{t} \geq k\} \cap \{\iota_{\omega}(\oslash) = \II\}) &=\sum_{n \in \N}\mathbb{P}(\{v_{t} \geq k\} \cap E(\II,k,n))\\
&=\sum_{n \in \N}\mathbb{P}(E(\II,k,n))\mathbb{P}(v_{t} \geq k \mid E(\II,k,n))\\
&=\sum_{n \in \N}p_\II(k,n)\beta_{t-1}^{n}\\ 
&=G_\II(k,\beta_{t-1}).
\end{align*}

It now follows that 
\[\beta_{t} = \mathbb{P}(v_{t} \geq k) = G_\I(k,0) + G_\I(k,1) - G_\I(k,\alpha_{t-1}) + G_\II(k,\beta_{t-1}),\]
and hence, recalling the definition of the vgf \eqref{eqn.vgf}, we obtain $\alpha_{t} = f_k(\alpha_{t-1})$. Unravelling this recursive relation yields $\alpha_{t} = f_k^{t}(\alpha_{0})$. Finally recall that $\omega_0$ is a trivial game, having a single node, $\oslash$. Consequently, $v_0 = \gamma_{\omega}(\oslash)$, and so $\alpha_{0} = \mathbb{P}(v_{0} < k) = p(\gamma < k) = f_{k}(0)$, where the last equality is by Lemma \ref{thm.vgf}(ii). Thus $\alpha_{t} = f_k^{t+1}(0)$.
\end{proof}

\subsection{The main results}\label{subsecn.main}
We are in a position to derive the main result of the paper. For $k > 0$ define \label{def.alpha}$\alpha(k) = \mathbb{P}(v < k)$ and \label{def.beta}$\beta(k) = \mathbb{P}(v \geq k)$. 

\begin{thm}\label{thm.fixed} 
Let $k > 0$. Then $\alpha(k)$ is the smallest fixed point of the function $f_{k}$.
\end{thm}
\begin{proof}
The result follows from Lemma \ref{thm.finite}: Let $s$ denote the smallest fixed point of $f_{k}$. Since $\alpha_{t+1} = f_{k}(\alpha_{t})$, and since the sequence $\alpha_{0}, \alpha_{1}, \ldots$ converges to $\alpha(k)$, continuity of $f_{k}$ implies that $\alpha(k) = f_{k}(\alpha(k))$, so that $\alpha(k)$ is a fixed point of $f_{k}$. Thus $s \leq \alpha(k)$. On the other hand, since $f_{k}$ is non-decreasing, we obtain by induction that $f_{k}^{t+1}(0) \leq s$. Hence by $\alpha_{t} \leq s$ for each $t \in \N$, and therefore $\alpha(k) \leq s$.
\end{proof}

\begin{thm}\label{thm.prob>0}
Let $k > 0$. Then $\beta(k) = 0$ if and only if all three of the following conditions are satisfied:
\begin{align}
p(\{\gamma < k\}) &> 0\label{eqn.cond1}\,,\\
p(\{\gamma \geq k\} \cap \{\xi =  0\}) &= 0\label{eqn.cond2}\,,\\
d(k) &\leq 1\label{eqn.cond3}\,.
\end{align}
\end{thm}
%Recall that $d(k) = \mathsf{E}_{p}(1_{\{\iota = \I\} \cap \{\gamma \geq k\}}\xi) + p(\{\iota = \II\} \cap \{\gamma \geq k\} \cap \{\xi = 1\}$.

Theorem \ref{thm.prob>0} follows directly from Theorem \ref{thm.fixed} and Lemma \ref{thm.vgf}.

That conditions \eqref{eqn.cond1} and \eqref{eqn.cond2} are necessary for $\beta(k) = 0$ follows easily from the following bounds:
\[p(\{\gamma \geq k\} \cap \{\xi =  0\}) \leq \beta(k) \leq 1 - p(\{\gamma < k\}).\]
The lower bound comes from the fact that in the event that the root of the tree has no children but a capacity of at least $k$, then also the value of the game is at least $k$. The upper bound holds since in the event that the value of the game is at least $k$, so is the capacity at the root of the tree. 

Condition \eqref{eqn.cond3} is much more subtle. Recall that $d(k)$, as defined in \eqref{eqn.d(k)}, is a sum of two terms, the expectation of the random variable $1_{\{\iota = \I\} \cap \{\gamma \geq k\}}\xi$, and the probability of the event $\{\iota = \II\} \cap \{\gamma \geq k\} \cap \{\xi = 1\}$. It is perhaps only natural that the first term affects \eqref{eqn.cond3}: intuitively, the higher is the expected number of nodes assigned to Player I with a capacity of at least $k$, the easier it is for Player I to secure a payoff of $k$. The second term is more difficult to interpret. Let us suggest one possible explanation: a player assigned to a node with a single child has no real choice of action at that node. Therefore, it is inconsequential who is being assigned to control the nodes having a single child: if one reassigns the nodes with a single child from Player II to Player I, one obtains a strategically equivalent game. 

We revisit condition \eqref{eqn.cond3} in the next section in the context of activation-independent escape models, where it is responsible for a phase transition with respect to Player I's activation probability. The discussion of $d(k)$ is further continued in Section \ref{secn.avoid}.

Theorem \ref{thm.prob>0} subsumes the classical criterion for the (non-)extinction of a branching process. To see this suppose that $\iota = \I$, that $\gamma = 1$ whenever $\xi \geq 1$, and that $\gamma = 0$ if $\xi = 0$. Then the value of a game $\omega$ is either $1$ or $0$, depending on whether the game tree $T_{\omega}$ has an infinite branch or not. Taking $k = 1$, Theorem \ref{thm.prob>0} reads $\mathbb{P}(v = 1) > 0$ if and only if $p(\{\xi = 0\}) = 0$ or $\mathsf{E}_{p}(\xi) > 1$. This can be easily seen to be equivalent to the classical condition: $p(\{\xi = 1\}) = 1$ or $\mathsf{E}_{p}(\xi) > 1$.

\section{Corollaries}\label{secn.cor}
We explore several features of the distribution of the value.   

Subsection \ref{subsecn.esssup} gives an expression for the essential supremum of the value.

When discussing the examples of Section \ref{secn.examples} we have noted that the probability of the event $\{v \geq k\}$ undergoes a phase transition with respect to the activation probability of Player I: it is positive only if Player I's activation probability is larger than a certain critical level. Subsection \ref{subsecn.critical} derives an expression for the $k$-critical level of Player I's activation probability for an activation-independent escape model.

Subsection \ref{subsecn.conditional} introduces into our study the distribution of the value conditional on the active player. In any activation-independent model, the distribution of the value at Player I's nodes first order stochastically dominates that at Player II's nodes, and both are non-decreasing (in the sense of first order stochastic dominance) with respect to Player I's activation probability. We also take a close look at the (conditional) probability of the event $\{v \geq k\}$ as Player I's activation probability approaches its $k$-critical value.

Subsection \ref{subsecn.asymptotic} discusses an asymptotic result for games defined on complete $n$-ary trees, as $n$ becomes large. 

Subsection \ref{subsecn.topology} looks at atoms in the distribution of the value. It also examines the distribution of the value as a function of the primitive distribution, and establishes sufficient conditions for this function to be continuous.  

\subsection{The essential supremum of the value}\label{subsecn.esssup} 
The essential supremum of the value is the quantity defined as 
\[\text{esssup } v = \inf\{k > 0:\mathbb{P}(v \geq k) = 0\}.\] 
It is the highest payoff Player I can be sure to get (provided she plays optimally) with positive probability. The following corollary is an immediate consequence of Theorem \ref{thm.prob>0} 

\begin{cor}
The essential supremum of the value is $\text{esssup }(v) = \max\{k_1,k_2,k_3\}$ where 
\begin{align*}
k_1 =& \inf\{k > 0:p(\{\gamma < k\}) > 0\}\,,\\
k_2 =& \inf\{k > 0:p(\{\gamma \geq k\} \cap \{\xi =  0\}) = 0\}\,,\\
k_3 =& \inf\{k > 0:d(k) \leq 1\}\,.
\end{align*}
\end{cor}

\subsection{Phase transitions in the escape model}\label{subsecn.critical} 
Since an escape model satisfies conditions \eqref{eqn.cond1} and \eqref{eqn.cond2} of Theorem \ref{thm.prob>0} for any $k > 0$, we obtain the following:   

\begin{cor}\label{thm.prob>0escape}
If $p$ is an escape model, then $\beta(k) = 0$ if and only if $d(k) \leq 1$.
\end{cor}

We have already noted an interesting feature of the two examples of Section \ref{secn.examples}: the probability of the event $\{v \geq k\}$ is only positive if Player I's activation probability is above a certain critical level (see Figures \ref{fig.fraclin} and \ref{fig.tree}). Here we give a general expression for the Player I's $k$-critical activation probability for an activation-independent escape model.

If $p$ satisfies activation-independence, $d(k)$ can be rewritten as
\[d(k) = q\cdot\mathsf{E}_{p}(1_{\{\gamma \geq k\}}\xi) + (1-q)\cdot p(\{\gamma \geq k\} \cap \{\xi = 1\}).\]
Solving the inequality $d(k)\le 1$ for $q$ we obtain $q \leq q_{c}(k)$, where $q_{c}(k)$, Player I's $k$--\textit{critical} activation probability, is defined by the following expression\label{def.q_c.b}:
\begin{equation}\label{eqn.criticalq}
q_{c}(k) = \begin{cases}0 & \text{if }\mathsf{E}_{p}(1_{\{\gamma \geq k\}}\xi) = \infty,\\\displaystyle\frac{1 - p(\{\gamma \geq k\} \cap \{\xi = 1\})}{\mathsf{E}_{p}(1_{\{\gamma \geq k\}}\xi) - p(\{\gamma \geq k\} \cap \{\xi = 1\})}&\text{if }1 < \mathsf{E}_{p}(1_{\{\gamma \geq k\}}\xi) < \infty,\\1 &\text{otherwise}.
\end{cases}
\end{equation}

We summarize the discussion as follows:

\begin{cor}
If $p$ is an activation-independent escape model, then $\beta(k) = 0$ if and only if $q \leq q_{c}(k)$, where $q_{c}(k)$ is given by \eqref{eqn.criticalq}  
\end{cor}

A further interpretation of the two terms appearing in \eqref{eqn.criticalq}, namely, $\mathsf{E}_{p}(1_{\{\gamma \geq k\}}\xi)$ and $p(\{\gamma \geq k\} \cap \{\xi = 1\})$ is offered by Corollary \ref{thm.ratiosbeta1}.

\subsection{Distribution of the value conditional on the active player}\label{subsecn.conditional} 
Apart from $\alpha(k)$ and $\beta(k)$, the probabilities of the events $\{v < k\}$ and $\{v \geq k\}$, one might also be interested in the probabilities of these events conditional on the root of the tree $\oslash$ being assigned to Player I or Player II. For $i \in \{\I,\II\}$, assuming that $q_i > 0$, let $\alpha_{i}(k) = \mathbb{P}(v < k \mid \iota(\oslash) = i)$ and $\beta_{i}(k) = \mathbb{P}(v \geq k \mid \iota(\oslash) = i)$. Of course, the conditional probabilities are related to the unconditional ones by $\alpha(k) = q_\I\alpha_\I(k) + q_\II\alpha_\II(k)$, and likewise for $\beta(k)$. 

For the rest of this section, fix a $k > 0$. We suppress the dependence on $k$ whenever convenient, writing e.g. $\alpha$, $\beta$, $\alpha_{i}$, $\beta_{i}$, and $q_{c}$, in place of $\alpha(k)$, etc.  

\begin{cor}\label{thm.betaIbetaII}
It holds that 
%\begin{align*}
%q_\I \cdot\beta_\I  &= \mathsf{E}_{p}(1_{\{\iota = \I\} \cap \{\gamma \geq k\}} (1_{\{\xi = 0\}} + 1 - \alpha^{\xi})),\\
%q_\II\cdot\beta_\II &= \mathsf{E}_{p}(1_{\{\iota = \II\} \cap \{\gamma \geq k\}}\beta^{\xi}).
%\end{align*}
\begin{align}
 q_{\I} \cdot\beta_{\I} &= G_{\I}(k,0) + G_{\I}(k,1) - G_{\I}(k,\alpha)\,,\\
 q_{\II}\cdot\beta_{II} &= G_{\II}(k,\beta)\,.
\end{align}
\end{cor}
\begin{proof}
Examining the proof of Lemma \ref{thm.finite} we find that
%\[\begin{aligned}
%&\mathbb{P}(\{v_{t} \geq k\} \cap \{\iota(\oslash) = \I\}) =& \mathsf{E}_{p}(1_{\{\iota = \I\} \cap \{\gamma \geq k\}} (1_{\{\xi = 0\}} + 1 - \alpha_{t-1}^{\xi}))\\
%&\mathbb{P}(\{v_{t} \geq k\} \cap \{\iota(\oslash) = \II\}) =& \mathsf{E}_{p}(1_{\{\iota = \II\} \cap \{\gamma \geq k\}}\beta_{t-1}^{\xi}).
%\end{aligned}\]
\begin{align}
\mathbb{P}(\{v_{t} \geq k\} \cap \{\iota(\oslash) = \I\}) &= G_{\I}(k,0) + G_{\I}(k,1) - G_{\I}(k,\alpha_{t-1})\,,\\
\mathbb{P}(\{v_{t} \geq k\} \cap \{\iota(\oslash) = \II\}) &= G_{\II}(k,\beta_{t-1})\,.
\end{align}
The result follows by taking the limit as $t \to \infty$ and using Lemma \ref{thm.finite}. 
\end{proof}

Recall that in an activation-independent model the random variables $(\gamma,\xi)$ and $\iota$ are independent. For such models we obtain two rather anticipated results: the value tends to be higher at nodes controlled by Player I than at nodes controlled by II. The statement is, of course, probabilistic: more precisely, the conditional distribution of the value given $\{\iota(\oslash) = \I\}$ first-order stochastically dominates that given $\{\iota(\oslash) = \II\}$. And secondly, the value is a ``non-decreasing" function of Player I's activation probability, again in the sense of first-order stochastic dominance. 

\begin{cor}\label{thm.ratiosbeta}
Suppose that $p$ is an activation-independent model. Then
\begin{itemize}
\item[(i)] It holds that 
\begin{align}
\beta_\I &= \mathsf{E}_{p}(1_{\{\gamma \geq k\}} (1_{\{\xi = 0\}} + 1 - \alpha^{\xi}))\label{eqn.betaI}\,,\\
\beta_\II &= \mathsf{E}_{p}(1_{\{\gamma \geq k\}}\beta^{\xi})\,.\label{eqn.betaII}
\end{align}
\item[(ii)] $\alpha_\I \leq \alpha \leq \alpha_\II$ and $\beta_\I \geq \beta \geq \beta_\II$. 
\item[(iii)] The probabilities $\alpha_\I$, $\alpha$, and $\alpha_\II$ are non-increasing while $\beta_\I$, $\beta$, and $\beta_\II$ are non-decreasing functions of Player I's activation probability $q$ on $[0,1]$.
\end{itemize}
\end{cor}
\begin{proof}
Let us define $G(x) = \mathsf{E}_{p}(1_{\{\gamma \geq k\}}x^{\xi})$. Then  
$G_{i}(k,x) = q_{i}G(x)$, from which Claim (i) follows by Corollary \ref{thm.betaIbetaII}.

For each $x \in [0,1]$ we have the inequality
\[
G(x) + G(1-x) - G(0) - G(1) = \mathsf{E}_{p}(1_{\{\gamma \geq k\}}(x^\xi + (1-x)^\xi - 1_{\{\xi = 0\}} - 1)) \leq 0.\]
In particular, $\beta_\II - \beta_\I = G(\beta) + G(\alpha) - G(0) - G(1) \leq 0$, from which Item (ii) follows.  

The vgf $f_{k}$ at a point $x \in [0,1]$ can be expressed as 
\begin{align*}
f_{k}(x) &= 1 - G_\I(k,0) - G_\I(k,1) + G_\I(k,x) - G_\II(k,1-x)\\ &= 1 - G(1-x) + q\left[G(x) + G(1-x) - G(0) - G(1)\right].
\end{align*}
This shows that $f_{k}(x)$ is a non-increasing function of $q$. Hence $\alpha$ is a non-increasing function of $q$, from which all the other assertions of Claim (iii) follow.
\end{proof}

We already know that at nodes controlled by Player I the probability of the event $\{v \geq k\}$ is larger than at nodes controlled Player II. How much larger is it? For an activation-independent escape model we compute the limits of the ratios $\beta_{i}/\beta$ as player I's activation probability $q$ approaches the $k$-critical level. The expressions reveal that, as Player I controls just enough nodes for the probability $\beta$ to be positive, $\beta_\I$ exceeds $\beta$ by the factor of $\mathsf{E}_{p}(1_{\{\gamma \geq k\}}\xi)$ while $\beta_\II$ is smaller than $\beta$ by the factor of $p(\{\gamma \geq k\} \cap \{\xi = 1\})$.

\begin{cor}\label{thm.ratiosbeta1}
Suppose that $p$ is an activation-independent escape model. Suppose moreover that \linebreak$\mathsf{E}_{p}(1_{\{\gamma \geq k\}}\xi) > 1$. Then
\begin{itemize}
\item[(iv)]  The ratio $\beta_\I/\beta$ is non-increasing and the ratio $\beta_\II/\beta$ is non-decreasing as functions of Player \I's activation probability $q$ on $(q_{c},1)$.
\item[(v)] As $q \downarrow q_{c}$, $\alpha$ approaches $1$ and $\beta$ approaches $0$. Moreover,
\begin{align}
\lim_{q \downarrow q_{c}}\frac{\beta_\I}{\beta} =& 
\mathsf{E}_{p}(1_{\{\gamma \geq k\}}\xi)\label{eqn.betaI/beta}\,,\\
\lim_{q \downarrow q_{c}}\frac{\beta_\II}{\beta} =& 
p(\{\gamma \geq k\} \cap \{\xi = 1\}).\label{eqn.betaII/beta}
\end{align}
\end{itemize}
\end{cor}
\begin{proof}
With the notation of Corollary \ref{thm.ratiosbeta} we have $G(0) = p(\{\gamma > k\} \cap \{\xi = 0\}) = 0$. Thus $\beta_\I =  G(1) - G(\alpha)$ and $\beta_\II = G(\beta)$. Claim (iv) follows by convexity of the function $G$.  

The first part of the Claim (v) follows since $\alpha = 1$ if $q = q_c$, and by Theorem \ref{thm.continuity}(i) below. As for the second part, 
\[\begin{aligned}
\lim_{q \downarrow q_{c}}\frac{\beta_\I}{\beta} =& \lim_{\beta \downarrow 0}\frac{G(1) - G(1-\beta)}{\beta} &&= G'(1) = \mathsf{E}_{p}(1_{\{\gamma \geq k\}}\xi)\,,\\
\lim_{q \downarrow q_{c}}\frac{\beta_\II}{\beta} =& \lim_{\beta \downarrow 0}\frac{G(\beta)}{\beta} &&= G'(0) = p(\{\gamma \geq k\} \cap \{\xi = 1\})\,,
\end{aligned}\]
by direct computation.
\end{proof}

\begin{exl}\label{exl.fraclin:cond}\rm
Let us start with a numerical illustration. Consider the setup of Example \ref{exl.fraclin} and let $l = 0.9$. The expected number of children is then $\mathsf{E}_{p}(\xi) = 9$, while the probability for a node to have a single child is $p(\xi = 1) = 0.09$. We may thus expect that, as $q$ approaches the 1--critical level $q_c = 0.1021$, the probability for Player I to win at the own node (i.e. at the node controlled by Player I) is approximately $100$ times larger than that at a node controlled by Player II. And indeed we find that for $q = 0.11$ the ratio $\beta_\I/\beta_\II$ is approximately $92$. 

We have 
\[\begin{aligned}
\alpha_\I &= \frac{1-l}{1-l\alpha}\,,&\quad&\beta_\I &= \frac{l\beta}{1-l+l\beta}\,,\\
\alpha_\II&= \frac{(1-l)^2 + (2l - l^2)\alpha}{1-l+l\alpha}\,,&&\beta_\II &= \frac{l(1-l)\beta}{1-l\beta}\,.
\end{aligned}\]
To derive $\beta_\I$ and $\beta_\II$ we use equations \eqref{eqn.betaI} and \eqref{eqn.betaII}, the fact that $\gamma = 0$ if $\xi = 0$, and the probability generating function of $\xi$: 
\[\mathsf{E}_{p}(x^{\xi}) = \frac{1-l}{1-lx}.\] The expressions for $\alpha_\I$ and $\alpha_\II$ are then obtained using $\alpha_{i} + \beta_{i} = 1$ and $\alpha + \beta = 1$.

The probabilities $\alpha$, $\alpha_\I$, and $\alpha_\II$ as functions of $q$ are pictured in the left panel of Figure \ref{fig.conditional}. In accordance with the preceding corollary, the $\alpha$ is the middle line (the same as that in Figure \ref{fig.fraclin}), $\alpha_\II$ is the top line, and $\alpha_\I$ is the bottom line. As $q$ approaches the $1$-critical level (see Example \ref{exl.fraclin} for an explicit expression), $\beta_\I/\beta$ converges to $l/(1-l)$, the expected number of children, while $\beta_\II/\beta$ converges to $l(1-l)$, the probability for a node to have exactly one child.

As $q$ approaches $1$ both $\alpha$ and $\alpha_\I$ converge to the same limit, the probability of ``extinction'' of the game tree (i.e. that the game tree is finite), $1/9$. On the other hand, $\alpha_\II$ converges to $3/5$. This is the probability that the root has no children, or that for at least one child of the root the corresponding subtree is finite. 
\end{exl}

\begin{exl}\label{exl.trees:cond}\rm
Consider now Example \ref{exl.tree}. We have:
\[\begin{aligned}
\alpha_\I &= 1 - (1-k) + (1-k)\alpha^n\,,&\quad&\beta_\I =& (1-k) - (1-k)(1-\beta)^n\,,\\
\alpha_\II&= 1 - (1-k)(1-\alpha)^n\,,&&\beta_\II =& (1-k)\beta^n\,.
\end{aligned}\]
These formulae follow easily from \eqref{eqn.betaI} and \eqref{eqn.betaII}. See the right panel of Figure \ref{fig.conditional} where $\alpha$, $\alpha_\I$, and $\alpha_\II$ are pictured for the ternary tree.

As $q$ approaches the $k$-critical level $q_c = 1/(1-k)n$, the ratio $\beta_\I/\beta$ approaches $(1-k)n$, while $\beta_\II/\beta$ approaches $0$. We interpret the latter fact as follows: if Player I's activation probability is just above the $k$-critical level, she can only guarantee a payoff of at least $k$ at her own nodes; the probability for Player I to secure a payoff of at least $k$ starting at a node controlled by Player II is negligible. %As $q$ approaches $1$, $\alpha$ and $\alpha_\I$ converge to the probability of ``extinction'' of the tree $T_{\omega}^{I}$ (see our discussion in the beginning of this section).

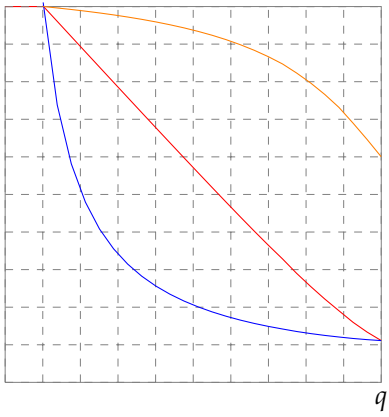
\begin{figure}
\caption{$\alpha$, $\alpha_\I$, and $\alpha_\II$ as functions of $q$.}\label{fig.conditional}
\begin{subfigure}{0.5\textwidth}
\begin{center}
\begin{tikzpicture}[domain=0:1,scale=5]
\draw[very thin,color=gray](0,0)--(1,0)--(1,1)--(0,1)--(0,0);
\draw[color=red](0,1)--(0.1,1);
\node[below] at(1,0){$q$};
\draw [step=0.1, gray, line width=0.2pt, dashed](0,0)grid(1,1);
\draw[color=red, domain=0.1:1] plot (\x,{0.5*(2-0.9)*(1-\x) + 0.5*(4*(0.1/0.9)^2 + ((2-0.9)*(1-\x))^2)^(0.5)});
\draw[color=blue, domain=0.101:1] plot (\x,{(1-0.9)/(1-0.9*(0.5*(2-0.9)*(1-\x) + 0.5*(4*(0.1/0.9)^2 + ((2-0.9)*(1-\x))^2)^(0.5))});
\draw[color=orange, domain=0.101:1] plot (\x,{((1-0.9)^2 + (2*0.9-0.9^2)*(0.5*(2-0.9)*(1-\x) + 0.5*(4*(0.1/0.9)^2 + ((2-0.9)*(1-\x))^2)^(0.5)))/(1-0.9+0.9*(0.5*(2-0.9)*(1-\x) + 0.5*(4*(0.1/0.9)^2 + ((2-0.9)*(1-\x))^2)^(0.5)))});
\end{tikzpicture}
\subcaption{Example \ref{exl.fraclin} with $l = 0.9$.}
\end{center}
\end{subfigure}
\begin{subfigure}{0.5\textwidth}
\begin{center}
\begin{tikzpicture}[domain=0:1,scale=5]
\draw[very thin,color=gray](0,0)--(1,0)--(1,1)--(0,1)--(0,0);
\draw[color=red, domain=0.3508:1] plot (\x,{0.5*(2 - 3*\x) + 0.5*((2 - 3*\x)^2 + 4*(1/19))^(0.5)});
\draw[color=blue, domain=0.3508:1] plot (\x,{1 - 0.95 + 0.95*(0.5*(2 - 3*\x) + 0.5*((2 - 3*\x)^2 + 4*(1/19))^(0.5))^3});
\draw[color=orange, domain=0.3508:1] plot (\x,{1 - 0.95*(1 - (0.5*(2 - 3*\x) + 0.5*((2 - 3*\x)^2 + 4*(1/19))^(0.5)))^3});
\node[below] at(1,0){$q$};
\draw [step=0.1, gray, line width=0.2pt, dashed](0,0)grid(1,1);
\end{tikzpicture}
\subcaption{Example \ref{exl.tree} with $n = 3$ and $k = 0.05$.}
\end{center}
\end{subfigure}
\end{figure}
\end{exl}

\subsection{An asymptotic result for games on complete $n$-ary trees}\label{subsecn.asymptotic}
We consider a special case of the model where (as in Example \ref{exl.tree}) the game tree is the complete $n$-ary tree. Fixing a particular joint distribution of the active player and the capacity, we study the probability of the event $\{v < k\}$ as $n$ becomes large. 

\begin{cor}\label{thm.asym}
Consider a sequence $p = p_{0},p_{1},\ldots$ of probability measures on $S$ such that the marginal of $p_{n}$ on $(\iota, \gamma)$ does not depend on $n$, and $p_{n}(\xi = n) = 1$. Let $k > 0$ be such that $0 < p(\{\gamma < k\})$. Denote $p(\{\gamma \geq k\} \cap \{\iota = i\})$ by $\rho_{i}$.
\begin{itemize}
\item The sequence $\mathbb{P}_{p_{n}}(v < k)$ is eventually monotone. It is:
\begin{itemize}
\item[{\rm (A)}] constant if $\rho_{\I} = 0$,  
\item[{\rm (B)}] eventually increasing if $\tfrac{1}{2} < \rho_{\I}$ and $0 < \rho_{\II}$, 
\item[{\rm (C)}] eventually decreasing otherwise. 
\end{itemize}
\item It converges to $1 - \rho_{\I}$. 
\end{itemize}
\end{cor}
\begin{proof} 
The vgf corresponding to $p_{n}$ is given by 
\begin{equation}\label{eqn.vgfn}
f_{n}(x) =  1 - \rho_{\I}(1 - x^{n}) - \rho_{\II}(1 - x)^{n}.
\end{equation}
Let us write $x_{n}$ for $\mathbb{P}_{p_{n}}(v < k)$. Recall that $x_{n}$ is the smallest fixed point of the function $f_{n}$.\smallskip

\noindent\textsc{Case A:} If $\rho_{\I} = 0$ then Theorem \ref{thm.prob>0} implies that $\mathbb{P}_{p_{n}}(v \geq k) = 0$ for each $n \in \N$, and we are done.\smallskip

For the rest of the proof we assume that $\rho_{\I} > 0$. Since $\rho_{\I} > 0$, condition \eqref{eqn.cond3} of Theorem \ref{thm.prob>0} is violated for $n \in \N$ sufficiently large, as the left-hand side of the inequality is $d(k) = n \rho_{\I}$. Consequently, $x_{n} < 1$ for $n \in \N$ sufficiently large. Moreover, it holds that $0 < x_{n}$ for each $n \in \N$ since $0 < p(\{\gamma < k\}) = f_n(0) \leq f_n(x_n) = x_n$. We find that 
\[0 < x_{n} < 1\text{ for }n \in \N\text{ sufficiently large}.\]
Below we use this fact together with Lemma \ref{thm.vgf}(iv) repeatedly.

Noting that $\rho_{\I} + \rho_{\II} = p(\{\gamma \geq k\}) < 1$, we can subdivide case C into three subcases as follows:
\begin{itemize}
\item[{\rm (C1)}] $0 < \rho_{\I}$ and $0 = \rho_{\II}$
\item[{\rm (C2)}] $0 < \rho_{\I} < \tfrac{1}{2}$ and $0 < \rho_{\II}$
\item[{\rm (C3)}] $\tfrac{1}{2} = \rho_{\I}$ and $0 < \rho_{\II}$.\smallskip
\end{itemize}

\noindent\textsc{Case C1:} Since $f_{n}(x) =  1 - \rho_{\I}(1 - x^{n})$ we find that $f_{n+1}(x_{n}) < f_{n}(x_{n}) = x_{n}$, implying that $x_{n+1} < x_{n}$. We conclude that the sequence $\{x_{n}\}_{n \in \N}$ is eventually decreasing. Hence $(x_{n})^{n} \to 0$. Taking the limit of $x_{n} = f_{n}(x_{n})$ we find that $x_{n} \to 1 - \rho_{\I}$.\smallskip  

For the rest of the proof we assume that $\rho_{\II} > 0$. Define 
\[y_{n} = 1 - \frac{1}{1 + \Big(\frac{\rho_{\II}}{\rho_{\I}}\Big)^{\frac{1}{n-1}}}.\]

As is easy to check, the equation $f_n(x)=f_{n+1}(x)$ admits exactly three solutions in $[0,1]$, namely $0$, $y_{n}$, and $1$. Moreover, 
\begin{align}
&f_n(x) < f_{n+1}(x)\text{ for }x \in (0,y_n)\label{eqn.asym1}\\
&f_{n+1}(x) < f_n(x)\text{ for }x \in (y_n,1).\label{eqn.asym2}
\end{align}
Indeed, this holds as at $x=0$ we have $f_{n}(0)=f_{n+1}(0)$ and $f_{n}'(0)<f_{n+1}'(0)$ and at $x=1$ we have $f_{n}(1)=f_{n+1}(1)$ and $f_{n}'(1)<f_{n+1}'(1)$. It also holds that
\begin{equation}\label{eqn.asym}
\lim_{n \to \infty}y_n = \tfrac{1}{2} \quad\text{and}\quad \lim_{n \to \infty}f_n(y_n) = 1-\rho_{\I}.
\end{equation}

\noindent\textsc{Case B:} By \eqref{eqn.asym}, $f_n(y_n) < y_{n}$ for large $n \in \N$. Hence $x_{n} < y_{n}$ for large $n \in \N$. It follows by \eqref{eqn.asym1} that $x_n = f_n(x_n) < f_{n+1}(x_n)$, hence $x_{n} < x_{n+1}$ for large $n \in \N$. We conclude that the sequence $\{x_{n}\}_{n \in \N}$ is eventually increasing. Since $x_{n} < y_{n}$ for large $n \in \N$, we have $\lim(x_{n})^{n} \leq \lim(y_{n})^{n} = 0$. And since $\{x_{n}\}_{n \in \N}$ is bounded away from zero, also $\lim(1 - x_{n})^{n} = 0$. Taking the limit of $x_{n} = f_{n}(x_{n})$ we find that $x_{n} \to 1 - \rho_{\I}$.\smallskip

\noindent\textsc{Case C2:} By \eqref{eqn.asym}, $y_n < f_n(y_n)$ for large $n \in \N$. Hence $y_n < x_n$ for large $n \in \N$. It follows by \eqref{eqn.asym2} that $f_{n+1}(x_n) < f_n(x_n) = x_n$, hence $x_{n+1} < x_n$ for large $n \in \N$. We conclude that the sequence $\{x_{n}\}_{n \in \N}$ is eventually decreasing. Hence $\lim(x_{n})^{n} = 0$. And since the sequence is bounded away from zero, also $\lim(1 - x_{n})^{n} = 0$. Taking the limit of $x_{n} = f_{n}(x_{n})$ we find that $x_{n} \to 1 - \rho_{\I}$.\smallskip

\noindent\textsc{Case C3:} Since $\rho_{\II} < \tfrac{1}{2} = \rho_{\I}$ we find that $\tfrac{1}{2} < f_{n}(\tfrac{1}{2})$, and hence $\tfrac{1}{2} < x_{n}$. On the other hand, $y_{n} < \tfrac{1}{2}$. We conclude that $y_{n} < x_{n}$ for all $n \in \N$. The rest of the argument is identical to that in the case C2.
\end{proof}

\begin{exl}\rm In the setup of Example \ref{exl.tree}, we have 
\[\lim_{n \to \infty} \mathbb{P}_{p_{n}}(v < k) = 1 - q + qk\]
for each $0 < k \leq 1$. The sequence $\mathbb{P}_{p_{n}}(v < k)$ is (A) constant if $q = 0$ or $k = 1$, (B) eventually increasing if $1/2 < q < 1$ and $0 < k < 1 - 1/2q$, and (C) eventually decreasing otherwise.
\end{exl}

\subsection{Continuity properties}\label{subsecn.topology}
We first look at atoms in the distribution of the value, i.e. points $k \geq 0$ such that $\PP(v = k) > 0$. As the following result shows, a point $k$ can only be an atom of the distribution of the value if it is either an atom or the essential infimum of the distribution of the capacity. The latter possibility is illustrated by Example \ref{exl.tree}, where the capacity has no atoms, but where the value could be zero almost surely. 

\begin{cor}
Let $k \geq 0$ be such that $p(\gamma = k) = 0$ and $p(\gamma < k) > 0$. Then $\PP(v = k) = 0$, and so $k$ is a continuity point of the value's cumulative distribution function $\alpha$.
\end{cor}
\begin{proof}
As is easy to see from the definition of the vgf \eqref{eqn.vgf}, $|f_{k_{0}}(x) - f_{k_{1}}(x)| < 4p(k_{0} \leq \gamma < k_{1})$ for any $x \in [0,1]$. In particular, the function $[0,\infty) \to [0,1]$, $k \mapsto f_{k}(x)$ is continuous.

Take a point $k \geq 0$ satisfying the hypothesis of the corollary and let $\bar{\alpha}(k) = \inf\{\alpha(k_1):k_1 > k\}$. Suppose that $\alpha(k) < \bar{\alpha}(k)$. Take any point $x$ with $\alpha(k) < x < \bar{\alpha}(k)$. Recall that $\alpha(k)$ is the smallest fixed point of the function $f_{k}$, and that $\alpha(k) > 0$ since $p(\gamma < k) > 0$. Lemma \ref{thm.vgf}(iv) now applies to show that $f_{k}(x) < x$. Take a $k < k_1$. It holds that $x < f_{k_{1}}(x)$. Indeed, we have $x < \alpha(k_1)$; If $\alpha(k_1) < 1$, then the desired inequality follows by Lemma \ref{thm.vgf}(iv), while if $\alpha(k_1) = 1$ it follows from the fact that $1$ is the only fixed point of $f_{k_{1}}$. 

Taking the limit $k_{1} \downarrow k$ we thus find that $f_{k}(x) < x \leq \lim_{k_{1} \downarrow k}f_{k_{1}}(x)$, contradicting the continuity of the function $k \mapsto f_{k}(x)$. 
\end{proof}

Recall that the measure $p$ is the primitive of our model. In what follows we view the cumulative distribution function of the value, $\alpha$, as a function of $p$. To make the dependence explicit we write $\alpha_{p}(k)$ to denote $\PP_{p}(v < k)$.

\begin{thm}\label{thm.continuity}
Let $\Delta_1$, respectively $\Delta_2$, denote the space of Borel probability measures on $S$, respectively on $[0,\infty)$, both endowed with the topology of weak convergence. 
\begin{itemize}
\item[(i)] Let $k > 0$. The function $\Delta_1 \to [0,1]$, $p \mapsto \alpha_{p}(k)$, is lower semicontinuous. It is continuous at a point $p \in \Delta_1$ if $p(\gamma = k) = 0$ and $p(\gamma < k) > 0$. 
\item[(ii)] The function $\alpha : \Delta_1 \to \Delta_2$ is continuous at a point $p \in \Delta_1$ if $p(\{\gamma < k\}) > 0$ for each $k > 0$. In particular, it is continuous at $p$ if $p$ is an escape model. 
\end{itemize}
\end{thm}
\begin{proof}
We write $f_k(x,p)$ for the values of the vgf to make the dependence on $p$ explicit. First we argue that for each $x \in [0,1]$, the function $f_k(x,\cdot) : \Delta_1 \to [0,1]$ is lower semicontinuous at each point of $\Delta_1$, and is continuous at a point $p \in \Delta_1$ such that $p(\gamma = k) = 0$. 

Write $f_k$ as $f_k(x,p) = 1 - \mathsf{E}_{p}(g)$. Here $g : S \to [0,1]$ is a function given by 
\[g(\gamma,\iota,\xi) = 1_{\{\gamma \geq k\}}[1_{\{\iota = \I\}} (1+1_{\{\xi=0\}}-x^{\xi}) + 1_{\{\iota = \II\}} (1-x)^{\xi}].\] 
The function $g$ is upper semicontinuous on $S$. This implies that the function $\Delta_1 \to [0,1]$ given by $p \mapsto \mathsf{E}_{p}(g)$ is upper semicontinuous. Moreover, $g$ is continuous on $S \setminus \{\gamma = k\}$. If $p(\gamma = k) = 0$, then $p$ is carried by the set of continuity points of $g$, and hence $p \mapsto \mathsf{E}_{p}(g)$ is continuous at the point $p$. 

For the rest of the proof fix a sequence $p_n$ in $\Delta_1$ converging weakly to $p$. Let
\[\underline{\alpha}(k) =: \liminf_{n \to \infty}\alpha_{p_{n}}(k)\text{ and }\bar{\alpha}(k) =: \liminf_{n \to \infty}\alpha_{p_{n}}(k).\]

Claim (i): For each $\epsilon > 0$:
\[\underline{\alpha}(k) =\liminf_{n \to \infty} f_{k}(\alpha_{p_{n}}(k),p_{n}) \geq \liminf_{n \to \infty} f_{k}(\underline{\alpha}(k) - \epsilon, p_{n}) \geq f_{k}(\underline{\alpha}(k) - \epsilon, p),\]
where the equation follows from the fact that $\alpha_{p_{n}}(k)$ is a fixed point of $f_{k}(\cdot,p_{n})$, the first inequality from monotonicity of $f_{k}(\cdot,p_{n})$, and the second inequality from lower semicontinuity of $f_{k}(\underline{\alpha}(k) - \epsilon, \cdot)$. Taking the limit as $\epsilon \downarrow 0$ we obtain $\underline{\alpha}(k) \geq f_{k}(\underline{\alpha}(k), p)$. As $0 \leq f_{k}(0,p)$, the intermediate value theorem implies that $f_{k}(\cdot,p)$ has a fixed point in $[0,\underline{\alpha}(k)]$. But since $\alpha_{p}(k)$ is the smallest fixed point of $f_{k}(\cdot,p)$, we conclude that $\alpha_{p}(k) \leq \underline{\alpha}(k)$. This proves the first part of the claim.

Let $p \in \Delta_1$ and $k > 0$ be such that $p(\gamma = k) = 0$ and $p(\gamma < k) > 0$. We show that $\bar{\alpha}(k) \leq \alpha_p(k)$. Suppose to the contrary and take any point $x$ such that $\alpha_{p}(k) < x < \bar{\alpha}(k)$. Since $0 < \alpha_{p}(k)$, Lemma \ref{thm.vgf}(iv) applies to show that $f_k(p,x) < x$. On the other hand, for infinitely many members of the sequence $\{p_n\}$ it holds that $x < \alpha_{p_{n}}(k)$. For each such member of the sequence, it holds that $x < f_k(x,p_n)$: indeed, if $\alpha_{p_{n}}(k) < 1$, the inequality is implied by Lemma \ref{thm.vgf}(iv), while if $\alpha_{p_{n}}(k) = 1$, it follows from the fact that $1$ is the only fixed point of $f_k(\cdot,p_{n})$. We thus find that $f_k(x,p) < x \leq \limsup_{n \to \infty}f_k(x,p_n)$, contradicting continuity of $f_k(x,\cdot)$ at $p$.

Claim (ii): Let $p \in \Delta_1$ be such that $p(\{\gamma < k\}) > 0$ for each $k > 0$. We argue that $\bar{\alpha}(k) \leq \alpha_p(k)$ whenever $k$ is a point of continuity of the cumulutive distribution function $\alpha_{p}$. Take an $\epsilon > 0$ and choose $k_1 > k$ so that $p(\gamma = k_1) = 0$ and $\alpha_p(k_1) \leq \alpha_p(k) + \epsilon$. Then $\bar{\alpha}(k) \leq \bar{\alpha}(k_1) \leq \alpha_p(k_1) \leq \alpha_p(k) + \epsilon$, where the middle inequality is by claim (i). Since $\epsilon > 0$ is arbitrary, we are done.
\end{proof}

Though it is not difficult to construct examples of discontinuity of the distribution of the value (one such example is below), these examples tend to be somewhat artificial. Indeed, we would assert that the distribution of the value is continuous at all primitive distributions of interest.

\begin{exl}\rm
Let $p$ be the Dirac measure on $(\iota,\gamma,\xi) = (\II,1,1)$, and let $p_n$ assign probability $1/n$ to the point $(\II,0,1)$ and probability $1 - 1/n$ to the point $(\II,1,1)$. Then $\PP_{p}(v = 1) = 1$ but $\PP_{p_n}(v = 0) = 1$ for each $n \in \N$. In particular, $\alpha : \Delta_{1} \to \Delta_{2}$ is not continuous at the point $p$.    
\end{exl}

\section{The conditional game}\label{secn.conditional}
Consider a game $\omega$ with a value $v_{\omega} \geq k$. In this section we study the subtree $T_{\omega}^*$ of the game tree $T_{\omega}$ consisting of those nodes where the value (of the corresponding subgame) is at least $k$. The tree $T_{\omega}^*$ characterizes Player I's $k$-optimal strategies, i.e. the strategies that allow her to ``defend" the payoff of $k$ against Player II: to guarantee the payoff of at least $k$, all Player I needs to do is never take an action leading outside the tree $T_{\omega}^*$. Another reason that we are interested in the tree $T_{\omega}^*$ is that it generalizes the so-called reduced family tree of a branching process: the latter consists of the individuals having an infinite line of descent (e.g. Athreya and Ney \cite[\S I.D.12]{Athreya}, Lyons and Peres \cite[\S 5.7]{LyonsPeres}).  

We define the ($k$-)conditional game, $\omega^*$, as the restriction of the game $\omega$ to the subtree $T_{\omega}^*$. As the above discussion makes it clear, the value of the game $\omega^*$ is the same as the value of $\omega$, provided that the latter has a value of at least $k$. The main technical result of the section states that conditional on the event $\{v_{\omega} \geq k\}$, the game $\omega^*$ has a distribution of a random perfect information game as in Section \ref{secn.model}. We identify the corresponding primitive distribution and compute the associated value generating function. 

We find a simple geometric relationship between the value generating function of the conditional game and the value generating function of the original game,  similar to the relationship that exists between the probability generating function of the reduced branching process and that of the original process  (Lyons and Peres \cite[Proposition 5.28(i)]{LyonsPeres}).

These results lead to a number of insights on the distribution of the conditional game, and in particular on the distribution of the tree $T_{\omega}^*$. Particularly interesting is the offspring distribution at Player I's nodes as it shows the number of moves that allow Player I to defend the value of $k$. 

The next subsection contains a formal definition of the conditional game and states the main result on its distribution. The second subsection highlights several features of the distribution of the conditional game. The third subsection illustrates these findings using examples. The final subsection contains the proofs.

The number $k > 0$ is fixed throughout this section. We write $\alpha = \alpha(k)$ and $\beta = \beta(k)$, and we assume that $\beta > 0$. 

\subsection{The definition and the main results}
%For a game $\omega \in \Omega$ with a value $v_{\omega} \geq k$ we define a subtree $T_{\omega}^*$ of the tree $T_{\omega}$ consisting of the nodes with a value of at least $k$, and the so-called upper conditional game $\omega^*$ in which the players' moves are restricted to the tree $T^*$. We establish that, conditionally on the event $v \geq k$, the game $\omega^*$ is follows the law of the  random perfect information game as is defined in Section 2, and identify the corresponding distribution. This analysis generalizes the textbook results on the reduced family tree of a branching process.

Consider a game $\omega \in \Omega$ with a value $v_{\omega} \geq k$. Define the tree $T_{\omega}^*$ to be the largest subtree of $T_{\omega}$ consisting of the nodes having a value of at least $k$. Equivalently one can define $T_{\omega}^*$ recursively as follows: the empty sequence $\oslash$ is an element of $T_{\omega}^*$. For $h \in T^*$ and $j \in \{1, \dots, \xi_{\omega}(h)\}$ declare $(h,j)$ to be an element of $T_{\omega}^*$ if $v_{\omega}(h,j) \geq k$.  

Intuitively, the tree $T_{\omega}^*$ consists of Player I's moves that allow him to ``defend'' a payoff of $k$ against Player II. To be able to guarantee the payoff of at least $k$, all Player I needs to do is never choose an action leading outside of the tree $T_{\omega}^*$. More formally, consider Player I's strategy $\sigma_\I$. Then $\sigma_\I$ is $k$-optimal if and only if whenever Player I's node $h$ is in $T_{\omega}^*$, $\sigma_\I$ selects a child $\sigma_\I(h) \in \{1, \dots, \xi_{\omega}(h)\}$ of $h$ such that the successor node $(h,\sigma_\I(h))$ is also in $T_{\omega}^*$. 

Recall that in an escape model, the capacity at any end node of the game tree is zero almost surely. In this case, the tree $T_{\omega}^{*}$ contains no end nodes, and its boundary $\partial T_{\omega}^{*}$ consists of the plays that are consistent with Player I's $k$-optimal strategies. 

In general, $T_{\omega}^{*}$ may well contain some of the end nodes of $T_{\omega}$. Consider a node $h \in T_{\omega}^*$ that is not an end node of $T_{\omega}$, that is, $\xi_{\omega}(h) \geq 1$. If $\iota(h) = \I$, then $h$ has a child, say $j \in \{1, \dots, \xi_{\omega}(h)\}$ such that $(h,j) \in T_{\omega}^*$; and if $\iota(h) = \II$, then $(h,j) \in T_{\omega}^*$ for all children $j \in \{1, \dots, \xi_{\omega}(h)\}$ of $h$. 

%In particular, if the tree $T^*$ is not empty, it is pruned. Also let $X^* \subseteq X$ denote the set of infinite branches of $T^*$. Then $X^*$ is not empty if and only if $T^*$ is not empty if and only if $v(\oslash) \geq k$. Moreover, the payoff of any play in $X^*$ is at least $k$. 

Define the $k$-\textit{conditional game} $\omega^*$\label{def.omega*} as a triple $(T_{\omega^{*}},\iota_{\omega^{*}},\gamma_{\omega^{*}})$, where $T_{\omega^{*}} = o(T_{\omega}^*)$ is the ordered copy of $T_{\omega}^*$, and $\iota_{\omega^{*}} = \iota_{\omega} \circ o^{-1}$ and $\gamma_{\omega^{*}} = \gamma_{\omega}  \circ o^{-1}$. Intuitively, $\omega^*$ is just a restriction of the game $\omega$ to the game tree $T_{\omega}^*$; our formalism, however, requires that $T_{\omega}^*$ be ordered.

Consider the value $v_{\omega^*}$ of the thus defined game $\omega^*$. Since $T_{\omega}^*$ only places restrictions on the moves of Player I, but not of Player II, we have $v_{\omega^*} \leq v_{\omega}$. On the other hand, by the remarks above, if $\sigma_\I$ is Player I's $k$-optimal strategy in $\omega$, then $\sigma_\I$ never requires Player I to leave the tree $T_{\omega}^*$, hence it is a legitimate strategy in $\omega^*$. This shows that $v_{\omega} \leq v_{\omega^*}$. We conclude that $v_{\omega} = v_{\omega^*}$.

The idea of the game $\omega^*$ generalizes the notion of the so-called reduced family tree of a branching process. Recall that the reduced family tree is a tree consisting of individuals having infinite line of descent. As is well known (e.g. Lyons and Peres \cite[Proposition 5.28(i)]{LyonsPeres}), conditional on non-extinction, the distribution of the reduced family tree is a Galton-Watson measure. The theorem below could be seen as providing a game-theoretic analogue to this classical result. 

\begin{lemma}
The map $\omega \mapsto \omega^*$, $\{\omega \in \Omega:v_{\omega} \geq k\} \to \Omega$ is Borel measurable.
\end{lemma}
The proof of the lemma is straighforwards and is omitted.

Define a primitive distribution $p^*$\label{def.p*} by letting $p_\I^*$ be given by
\[\begin{aligned}
p_\I^*(c,0) &= p_\I(c,0)\beta^{-1},& c \geq k,\\ p_\I^*(c,n) &= \sum_{m \geq n} p_\I(c,m) \binom{m}{n} \beta^{n-1} \alpha^{m-n},& c \geq k, n \geq 1,\\ p_\I^*(c,n) &= p_\I^*(k,n),& c < k,  n\geq 0,
\end{aligned}\]
and letting $p_\II^*$ be given by
\[\begin{aligned}
p_\II^*(c,0) &= p_\II(c,0)\beta^{-1},& c \geq k,\\ p_\II^*(c,n) &= p_\II(c,n)\beta^{n - 1},& c \geq k, n \geq 1,\\ p_\II^*(c,n) &= p_\II^*(k,n),& c < k, n \geq 0.
\end{aligned}\]

The measure $p^*$ governs the distribution of the conditional game, in the sense made precise by Theorem \ref{thm.upperconditional}. We offer some intuition behind these expressions. 

The bottom equation of both triples (for the case $c < k, n \geq 0$) simply say that the capacity is at least $k$ with $p^*$-probability $1$. This is in accordance with our definitions: a node of $\omega$ cannot be included in the game tree of $\omega^*$ unless it has a capacity of at least $k$. 

The top equations of the triples (for the case $c \geq k$) are driven by the fact that a node of the tree $\omega^*$ has no children in $\omega^*$ only if it has no children at all. Thus $p_{i}^*(c,0)$ is the probability that the node has the active player $i$, the capacity of at least $c$, and no children, conditional on the event that the value at the node is at least $k$. 

The core of the definition are the two middle equations (for the case $c \geq k, n \geq 1$). The one corresponding to Player I describes the probability that a node has I as the active player, the capacity of at least $c$, and exactly $n$ children with a value of at least $k$, given that the node itself has a value of at least $k$. The index $m$ represents the total number of children of the node, and the expression under the summation sign is the probability of exactly $n$ successes out of $m$ independent experiments, with $\beta$ being the probability of success; the expression is divided by $\beta$ as we condition on the event of the node having a value of at least $k$. The formula corresponding to Player II reflects the fact that, if Player II's node is included in the game tree of $\omega^*$, all of its children are as well.  

Our main result on the conditional game is the following theorem. The statement is quite intuitive given the defining formulas for $p^*$ and thus the somewhat technical proof is postponed to Section \ref{secn.law_of_p*}.

\begin{thm}\label{thm.upperconditional}
Suppose that $\beta > 0$. Conditional on the event $\{\omega \in \Omega:v_{\omega} \geq k\}$, the game $\omega^*$ is distributed according to the measure $\mathbb{P}_{p^*}$: that is to say $\mathbb{P}(\{\omega^* \in B \} \mid \{v_{\omega} \geq k\}) = \mathbb{P}_{p^*}(B)$ for each Borel set $B \subseteq \Omega$.
\end{thm}

We presently state two key consequences of the theorem, and leave a more detailed exploration of the properties of $p^*$ to the next subsection. The first is the expression for the value generating function (denoted $f^*$\label{def.f*}) associated to the primitive distribution $p^*$. The second is the fact that $T_{\omega}^*$ is a Galton-Watson tree.

\begin{cor}\label{thm.vgf*}
For $c \geq k$ and $x \in [0,1]$ we have:
\begin{equation}\label{eqn.vgf*}
f^*(c,x) = \frac{f(c,\beta x +\alpha) - \alpha}{\beta}.
\end{equation}
\end{cor}

\begin{figure}[h]
\begin{center}
\begin{tikzpicture}[domain=0:1,scale=5]
\draw [step=0.1, gray, line width=0.2pt, dashed](0,0)--(1,1);
\draw[very thin,color=gray](0,0)--(1,0)--(1,1)--(0,1)--(0,0);
\draw[thick,color=black](0.3028,0.3028)--(1,0.3028)--(1,1)--(0.3028,1)--(0.3028,0.3028);
\draw[color=orange, domain=0:1] plot (\x,{1 - (1-0.30)*0.7 + (1-0.30)*0.7*\x^6 - (1-0.30)*(1-0.7)*(1-\x)^6)});
\draw[color=red, domain=0:1] plot (\x,{1 - (1-0.05)*0.7 + (1-0.05)*0.7*\x^6 - (1-0.05)*(1-0.7)*(1-\x)^6)});
%\node[below] at(0,0){0};
%\node[below] at(1,0){1};
%\node[left] at(0,0){0};
%\node[left] at(0,1){1};
\end{tikzpicture}
\caption{The functions $f_{0.05}$ (red) and $f_{0.30}$ (orange) in Example \ref{exl.tree} with $n = 6$ and $q = 0.7$. The portions of the graphs within the bold black square represent, up to an appropriate rescaling, the graphs of the value generating functions $f_{0.05}^*$ and $f_{0.30}^*$ in the $0.05$-conditional game.}
\label{fig.vgfcond}
\end{center}
\end{figure}
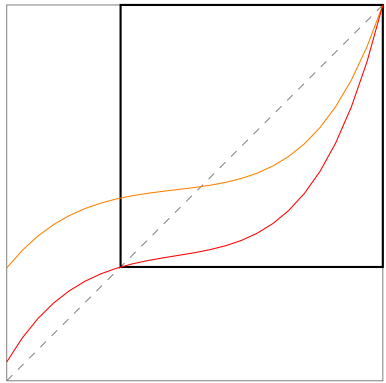

Formula \eqref{eqn.vgf*} reveals a simple geometric relationship between the vgf of $p$ and that of $p^*$. To obtain the graph of $f_{c}^*$, one stretches the graph of $f_{c}$ so that the point $(\alpha,\alpha)$ becomes aligned with the origin. Figure \ref{fig.vgfcond} illustrates. In particular, $x^*$ is a fixed point of $f_{c}^*$ exactly when $\beta x^* + \alpha$ is a fixed point of $f_{c}$. The point $0$ is a fixed point of $f_{k}^*$. Both observations, of course, are a could also be deduced from the fact that $\omega^*$ has the same value as $\omega$. 

We think of formula \eqref{eqn.vgf*} as a game-theoretic analogue of the textbook result on the reduced family tree of the branching processes (Lyons and Peres \cite[Proposition 5.28(i)]{LyonsPeres}). The reduced family tree of a branching process is distributed according to a Galton-Watson measure, i.e. it is itself a family tree of a branching process. Letting $k = c = 1$ in \eqref{eqn.vgf*} one obtains the pgf of the corresponding offspring distribution in terms of the pgf of the offspring distribution of the original process.

\begin{cor}\label{thm.GW*}
Suppose that $\beta > 0$. Conditional on the event $\{\omega \in \Omega:v_{\omega} \geq k\}$, the distribution of the tree $T_{\omega^*}$ is the Galton-Watson measure generated by an offspring distribution with mean 
\begin{equation}\label{eqn.Ep*(xi)}
\mathsf{E}_{p^*}(\xi) = \mathsf{E}_{p}(1_{\{\iota = \I\} \cap \{\gamma \geq k\}}\xi + 1_{\{\iota = \II\} \cap \{\gamma \geq k\}} \xi\beta^{\xi-1}).
\end{equation}
It holds that $d(k) \leq \mathsf{E}_{p^*}(\xi)$.
\end{cor}

Both corollaries follow easily from Lemma \ref{thm.p*}(i) below.

\subsection{Properties of the distribution of the conditional game}
We list some of the main properties of the measure $p^*$. Note that item (i) of Lemma \ref{thm.p*} completely characterizes the measure $p^*$, and can be taken to be its definition. The function $G_i$ is defined by \eqref{eqn.G} with $p$ replaced by $p^*$.

\begin{lemma}\label{thm.p*} 
Suppose that $\beta = \beta(k) > 0$.
\begin{itemize} 
\item[(i)] For $c \geq k$ and $x \in [0,1]$:
\begin{align*}
G_\I^*(c,x)  &= \frac{G_\I(c,\alpha + \beta x) + G_\I(c,0) - G_\I(c, \alpha)}{\beta},\\
G_\II^*(c,x) &= \frac{G_\II(c,\beta x)}{\beta}.
\end{align*}
\item[(ii)] Under the measure $p^*$, the capacity is at least $k$ almost surely. 
\item[(iii)] The activation probability under the measure $p^*$ are given, for $i \in \{\I,\II\}$ by 
\begin{equation}\label{eqn.p*(i)}
p^*(\iota = i) = \frac{\beta_{i}}{\beta}p(\iota = i).
\end{equation}
\item[(iv)] 
\begin{align*}
\mathsf{E}_{p^*}(1_{\{\iota = \I\}} \xi) &= \mathsf{E}_{p}(1_{\{\iota = \I\} \cap \{\gamma \geq k\}}\xi)\,,\\ 
\mathsf{E}_{p^*}(1_{\{\iota = \II\}}\xi) &= \mathsf{E}_{p}(1_{\{\iota = \II\} \cap \{\gamma \geq k\}} \xi\beta^{\xi-1}).
\end{align*}
\end{itemize}
\end{lemma}
\begin{proof}
Claim (i) is established by a direct computation, while Claim (ii) is immediate from the definition of $p^*$. Turning to Claims (iii) and (iv) we note that, as a consequence of Claim (ii), for $x \in [0,1]$
\[\mathsf{E}_{p^*}(1_{\{\iota = i\}} x^\xi) = G_{i}^*(k,x).\]
In particuar, $p^*(\iota = i) = G_{i}^*(k,1)$. Using Corollary \ref{thm.betaIbetaII} we compute: 
\[G_\I^*(k,1) = \tfrac{1}{\beta}[G_\I(k,0) + G_\I(k,1) - G_\I(k,\alpha)] = \tfrac{q_\I \cdot \beta_\I}{\beta}\,,\] 
\[G_\II^*(k,1) = \tfrac{1}{\beta}G_\II(k,\beta) = \tfrac{q_\II \cdot \beta_\II}{\beta}.\]
Finally, 
\begin{align*}
\mathsf{E}_{p^*}(1_{\{\iota = \I\}} \xi) &= \frac{\partial}{\partial x}G_\I^*(k,1) = \frac{\partial}{\partial x}G_\I(k,1) =  \mathsf{E}_{p}(1_{\{\iota = \I\} \cap \{\gamma \geq k\}} \xi),\\
\mathsf{E}_{p^*}(1_{\{\iota = \II\}} \xi) &= \frac{\partial}{\partial x}G_\II^*(k,1) = \frac{\partial}{\partial x}G_\II(k,\beta) = \mathsf{E}_{p}(1_{\{\iota = \II\} \cap \{\gamma \geq k\}} \xi\beta^{\xi-1}).
\end{align*}
\end{proof}

%Claim (ii) of Lemma \ref{thm.p*} reflects the most basic property of the conditional game: each node of its game tree has a capacity of at least $k$, and at least one child. Claim (iii) is arguably very natural as well, and claim (iv) points to the asymmetry in the distribution of the nodes at Player I and at Player II's nodes.

Perhaps not surprisingly, $p^*$ need not be an activation-independent model, even if $p$ is: indeed, in the conditional game the offspring distribution at Player I's nodes might well be different from that at Player II's nodes. This feature of the conditional game is reflected in the corollary below, and is further illustrated by the examples that follow.

\begin{cor}
Suppose that $p$ is an activation-independent model. Then
\begin{itemize}
\item[(i)] Player I is assigned to control a node more frequently under the measure $p^*$ than under $p$: $p^*(\iota = \I) \geq p(\iota = \I)$.
\item[(ii)] It holds that
\begin{align}
\mathsf{E}_{p^*}(\xi \mid \iota = \I) &= \frac{\beta}{\beta_\I}\mathsf{E}_{p}(1_{\{\gamma \geq k\}}\xi)\label{eqn.xi*I}\\
\mathsf{E}_{p^*}(\xi \mid \iota = \II)&= \frac{\beta}{\beta_\II} \mathsf{E}_{p}(1_{\{\gamma \geq k\}}\xi\beta^{\xi-1}).\label{eqn.xi*II}
\end{align}
In particular, for both $i = \I$ and $i = \II$ we have 
\[\mathsf{E}_{p^*}(\xi \mid \iota = i) \leq \mathsf{E}_{p}(1_{\{\gamma \geq k\}}\xi).\]
\item[(iii)] Suppose furthermore that $p$ is an escape model and that $1 < \mathsf{E}_{p}(1_{\{\gamma \geq k\}}\xi) < \infty$. Then
\begin{itemize}
\item[(a)] $p^*(\xi = 0) = 0$. 
\item[(b)] $\mathsf{E}_{p^*}(\xi \mid \iota = \I)$ is a non-decreasing function of $q$ on $(q_{c},1)$.
\item[(c)] It holds that
\begin{align*}
&\lim_{q \downarrow q_{c}} p^*(\iota = \I) = 
\frac{\mathsf{E}_{p}(1_{\{\gamma \geq k\}}\xi) [1 - p(\{\gamma \geq k\} \cap \{\xi = 1\})]}{\mathsf{E}_{p}(1_{\{\gamma \geq k\}}\xi) - p(\{\gamma \geq k\} \cap \{\xi = 1\})}\,,\\
%\lim_{q \downarrow q_{c}} p^*(\iota = \II) &= p(\xi = 1) \frac{p(\gamma \geq k) \mathsf{E}_{p}(\xi) - 1}{\mathsf{E}_{p}(\xi) - p(\xi = 1)}\\
&\lim_{q \downarrow q_{c}} p^*(\xi  = 1) = 1\,,\\
&\lim_{q \downarrow q_{c}} p^*(\xi  = 1\mid \iota = \I) = 1\,,\\
&\lim_{q \downarrow q_{c}} p^*(\xi  = n\mid \iota = \II) = 1\,,
\end{align*}
where $n$ is the smallest number in $\N_{+}$ such that $p(\{\gamma \geq k\} \cap \{\xi = n\}) > 0$.
\end{itemize}
\end{itemize}
\end{cor}

\begin{proof}
Claims (i) and (ii) follow at once from Lemma \ref{thm.p*} and Corollary \ref{thm.ratiosbeta}. Only Claim (iii) warrants some explanation. 

Item (a) follows directly from the definition of $p^*$ and the assumption that $p$ is an escape model. Item (b) follows from Claim (ii) and Corollary \ref{thm.ratiosbeta1}. Consider Item (c).

The first equation: using equations \eqref{eqn.p*(i)} and \eqref{eqn.betaI/beta} we obtain 
\[\lim_{q \downarrow q_{c}} p^*(\iota  = \I) = \lim_{q \downarrow q_{c}} \frac{\beta_{I}}{\beta}q = \mathsf{E}_{p}(1_{\{\gamma \geq k\}}\xi) \cdot q_{c}.\] Finally, use the definition of $q_{c}$ in \eqref{eqn.criticalq}.

The second equation: We already know that $\xi \geq 1$ almost $p^{*}$-surely. Hence $2- \mathsf{E}_{p^*}(\xi) \leq p^{*}(\xi = 1)$. Therefore it suffices to argue that $\mathsf{E}_{p^*}(\xi)$ converges to $1$. Using equation \eqref{eqn.Ep*(xi)}, the assumption that $\iota$ is independent of $(\gamma, \xi)$, the fact that $\beta$ converges to $0$ as $q \downarrow q_{c}$ (see Corollary \ref{thm.ratiosbeta1}), and the formula \eqref{eqn.criticalq}, we obtain:
\begin{align*}
\lim_{q \downarrow q_{c}}\mathsf{E}_{p^*}(\xi) 
&= \lim_{q \downarrow q_{c}} q \cdot \mathsf{E}_{p}(1_{\{\gamma \geq k\}}\xi) + \lim_{q \downarrow q_{c}} (1 - q) \cdot \mathsf{E}_{p}(1_{\{\gamma \geq k\}} \xi\beta^{\xi-1})\\
&= q_{c} \cdot \mathsf{E}_{p}(1_{\{\gamma \geq k\}}\xi) + (1-q_{c}) \cdot p(\{\gamma \geq k\} \cap \{\xi = 1\}) = 1.
\end{align*}

The third equation: By a similar reasoning, it suffices to argue that $\mathsf{E}_{p^*}(\xi \mid \iota = \I)$ converges to $1$. The latter fact follows directly from equation \eqref{eqn.xi*I} and \eqref{eqn.betaI/beta}. 

The fourth equation: using the definition of $p^{*}$ and equation \eqref{eqn.betaII} we obtain
\begin{align*}
p^*(\xi  = n \mid \iota = \II)  &= \frac{p_\II^*(0,n)}{p^{*}(\iota = \II)}\\ &= \frac{p_\II(k,n) \beta^{n-1}}{p(\iota = \II)} \frac{\beta}{\beta_\II}\\ & = \frac{p(\{\iota = \II\} \cap \{\gamma \geq k\} \cap \{\xi = n\}) }{p(\iota = \II)} \cdot \frac{\beta^{n}}{\mathsf{E}_{p}(1_{\{\gamma \geq k\}}\beta^{\xi})}\\ & =  \frac{p(\{\gamma \geq k\} \cap \{\xi = n\})}{\mathsf{E}_{p}(1_{\{\gamma \geq k\}}\beta^{\xi - n})}.
\end{align*}
As $q \downarrow q_{c}$, $\beta$ converges to $0$, and the expression on the extreme right-hand side of the above array converges to $1$.
\end{proof}

Claim (i) above is arguably very intuitive. As the nodes controlled by Player I are more likely to have a value of at least $k$ than the nodes controlled by Player II, in the conditional game Player I is assigned to be an active player more frequently than in the original game.

Claim (ii) points to the fact that, in general, $p^*$ is not activation-independent even if $p$ is: the offspring distribution of the conditional game at Player I's node differs from that at Player II's nodes. The two examples below provide an illustration.

We comment on Claim (iii). The mean of the offspring distribution at Player I's nodes in the conditional game is non-decreasing in $q$, the activation probability under $p$. Thus the higher the activation probability, the higher is, in expectation, the average number of moves that Player I could make to defend the value of $k$ against Player II. Moreover, if Player I's activation probability is close to its $k$--critical level $q_{c}(k)$, ``typically'' only one such move is available to Player I. 

The mean of the offspring distribution at Player II's nodes in the conditional game is not, in general, monotone in $q$ (see Example \ref{exl.treesupper}). As $q$ converges to the $k$--critical level, the offspring distribution at Player II's nodes becomes degenerate. If $p(\{\gamma \geq k\} \cap \{\xi = 1\}) > 0$, it converges to a pointmass on $1$ (as is the case in Example \ref{exl.linfracupper}); otherwise, it converges to a pointmass on $n$, the smallest natural number with $p(\{\gamma \geq k\} \cap \{\xi = n\}) > 0$ (as is the case in Example \ref{exl.treesupper}).  

Moreover, the mean of the offspring distribution at Player I's nodes could be lower as well as higher than that at Player II's nodes, another point highlighted by the examples below.   
\subsection{Examples}
\begin{exl}\label{exl.linfracupper}\rm
Let us first give a numerical illustration. Consider Example \ref{exl.fraclin} with $l = 0.9$. The expected number of children under $p$ is then $\mathsf{E}_{p}(\xi) = 9$. First let $q = 0.11$, just above the 1--critical level (0.1021). Then, under the measure $p^*$, player I is assigned to a node with probability $p^*(\iota = \I) = 0.9193$. The expected number of children of a node controlled by Player I is $\mathsf{E}_{p^*}(\xi \mid \iota = \I) = 1.0769$, and that at a node controlled by Player II is $\mathsf{E}_{p^*}(\xi \mid \iota = \I) = 1.0077$. Both numbers are ``barely'' above $1$, indicating that a node of the conditional game has exactly 1 child with a very high probability. 

Letting $q = 0.9$ player I's activation probability under $p^*$ increases to $0.9655$. The expected number of children of a node controlled by Player I and Player II, respectively, both increase to $8.39$ and $3.83$, respectively. 

Under the measure $p^*$, player I is chosen with probability
\[p^*(\iota = \I) = \frac{l}{1-l+l\beta}q\]
(recall that $\beta=\beta(k)$).
%\[\begin{aligned}
%p^*(\iota = \I) &= \frac{l}{1-l+l\beta}q_\I\\
%p^*(\iota = \II) &= \frac{l(1-l)}{1-l\beta}q_\II.
%\end{aligned}\]
The expected number of children, conditional on the active player, are given by
\[\mathsf{E}_{p^*}(\xi \mid \iota = \I) = 1 + \frac{l}{1-l}\beta\,\text{ and }\,\mathsf{E}_{p^*}(\xi \mid \iota = \II) = \frac{1}{1-l\beta}.\]
To obtain these expressions, use \eqref{eqn.p*(i)}, \eqref{eqn.xi*I}, and \eqref{eqn.xi*II}, the expressions for $\beta_{\I}$ and $\beta_{\II}$ from Example \ref{exl.fraclin:cond}, and the following fact: 
\[\mathsf{E}_{p}(\xi x^{\xi-1}) = \frac{\partial}{\partial x} \mathsf{E}_{p}(x^{\xi}) = \frac{\partial}{\partial x} \frac{1-l}{1-lx} = \frac{l(1-l)}{(1-lx)^2}.\] 
One can see that $\mathsf{E}_{p^*}(\xi \mid \iota = \II) \leq \mathsf{E}_{p^*}(\xi \mid \iota = \I)$ (the two expressions are equal to each other if $\beta = 0$ or if $\beta = 1$; the former is convex, while the latter is linear in $\beta$). The unconditional expected number of children under $p^*$ is
\[\mathsf{E}_{p^*}(\xi) = q_\I\frac{l}{1-l} + q_\II\frac{l(1-l)}{(1 - l\beta)^2}.\]
\begin{figure}[h]
\caption{The conditional game in Example \ref{exl.fraclin} with $l = 0.9$}
\label{fig.fraclin:uppercond}
\begin{subfigure}{.5\textwidth}
\begin{center}
\begin{tikzpicture}[domain=0:1,xscale=5,yscale=50,variable=\q]
\draw[very thin,color=gray](0,0.9)--(1,0.9)--(1,1)--(0,1)--(0,0.9);
\draw[color=blue, domain=0.11:1] plot (\q,{0.9*\q/(1 - 0.9*(0.5*(2-0.9)*(1-\q) + 0.5*(4*(0.1/0.9)^2 + ((2-0.9)*(1-\q))^2)^(0.5)))});
\node[below] at(1,0.9){$q$};
\node[left] at(0,1){1};
\node[left] at(0,0.95){0.95};
\node[left] at(0,0.9){0.9};
%\draw[color=red, domain=0.1:1] plot (\x,{0.5*(2-0.9)*(1-\x) + 0.5*(4*(0.1/0.9)^2 + ((2-0.9)*(1-\x))^2)^(0.5)});
\draw [xstep=0.1, ystep=0.01, gray, line width=0.2pt, dashed](0,0.9)grid(1,1);
\end{tikzpicture}
\end{center}
\subcaption{$p^*(\iota = \I)$ as a function of $q$  (note the scale of the vertical axis).}
\end{subfigure}\hspace{1cm}
\begin{subfigure}{.5\textwidth}
\begin{center}
\begin{tikzpicture}[domain=0:1,xscale=5,yscale=0.625,variable=\q]
\draw[very thin,color=gray](0,1)--(1,1)--(1,9)--(0,9)--(0,1);
\draw[color=blue, domain=0.11:1] plot (\q,{(1 - 0.9*(0.5*(2-0.9)*(1-\q) + 0.5*(4*(0.1/0.9)^2 + ((2-0.9)*(1-\q))^2)^(0.5)))/0.1});
\draw[color=red, domain=0.11:1] plot (\q,{1/(1-0.9+0.9*(0.5*(2-0.9)*(1-\q) + 0.5*(4*(0.1/0.9)^2 + ((2-0.9)*(1-\q))^2)^(0.5)))});
\node[below] at(1,1){$q$};
\node[left] at(0,1){$1$};
\node[left] at(0,9){$9$};
\node[left] at(0,5){$5$};
\draw [xstep=0.1, ystep=1, gray, line width=0.2pt, dashed](0,1)grid(1,9);
\end{tikzpicture}
\end{center}
\subcaption{$\mathsf{E}_{p^*}(\xi \mid \iota = \I)$ (blue) and $\mathsf{E}_{p^*}(\xi \mid \iota = \II)$ (red) as functions of $q$.}
\end{subfigure}
\end{figure}
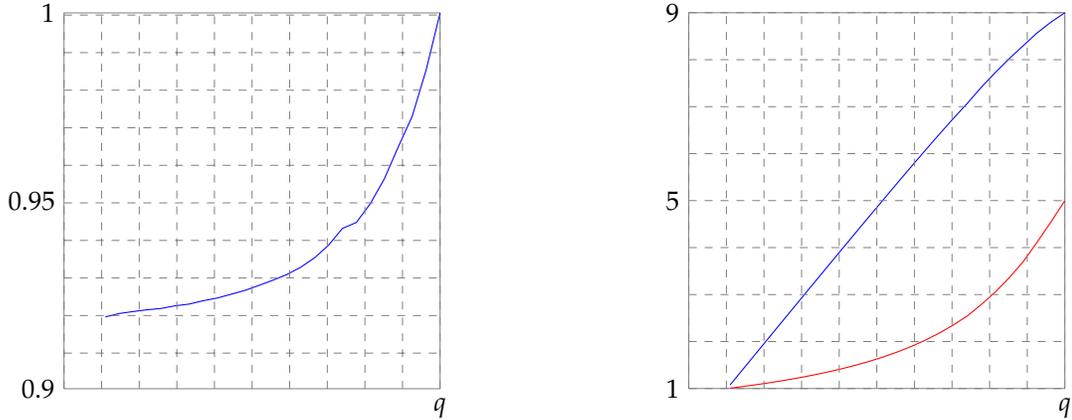
\end{exl}

\begin{exl}\label{exl.treesupper}\rm
Consider Example \ref{exl.tree}. Under the measure $p^*$ the activation probability of Player I is (this and the other formulae of the example follow by combining equations \eqref{eqn.p*(i)}, \eqref{eqn.xi*I}, and \eqref{eqn.xi*II}, and those in Example \ref{exl.trees:cond})  
\[p^*(\iota = \I) = \frac{(1-k) - (1-k)(1-\beta)^n}{\beta}q.\]
Somewhat surprisingly, it is not monotone in $q$. The left panel of Figure \ref{fig.tree:upper} illustrates for $n = 3$ and $k = 0.05$. Player I's activation probability $p^*(\iota = \I)$ under $p^*$ approaches $1$ if $q \downarrow q_{c}(k)$ and if $q \uparrow 1$, while for $q = 0.7$, we find that $p^*(\iota = \I) = 0.8106$. 

\begin{figure}[h!]
\caption{The conditional game in Example \ref{exl.tree} with $n = 3$ and $k = 0.05$.}
\label{fig.tree:upper}
\begin{subfigure}{.5\textwidth}
\begin{center}
\begin{tikzpicture}[domain=0:1,scale=5]
\draw[very thin,color=gray](0,0)--(1,0)--(1,1)--(0,1)--(0,0);
\draw[color=blue, domain=0.3511:1] plot (\x,{\x*(1-0.05)*(1-(0.5*(2 - 3*\x) + 0.5*((2 - 3*\x)^2 + 4*(1/19))^(0.5))^3)/(1-(0.5*(2 - 3*\x) + 0.5*((2 - 3*\x)^2 + 4*(1/19))^(0.5)))});
\node[below] at(1,0){$q$};
\node[left] at(0,0){$0$};
\node[left] at(0,1){$1$};
\draw [step=0.1, gray, line width=0.2pt, dashed](0,0)grid(1,1);
\end{tikzpicture}
\end{center}
\subcaption{$p^*(\iota = \I)$ as a function of $q$.}
\end{subfigure}
\begin{subfigure}{.5\textwidth}
\begin{center}
\begin{tikzpicture}[domain=0:1,xscale=5,yscale=2.5]
\draw[very thin,color=gray](0,1)--(1,1)--(1,3)--(0,3)--(0,1);
\draw[color=blue, domain=0.3511:1] plot (\x,{3/(1+(0.5*(2 - 3*\x) + 0.5*((2 - 3*\x)^2 + 4*(1/19))^(0.5))+(0.5*(2 - 3*\x) + 0.5*((2 - 3*\x)^2 + 4*(1/19))^(0.5))^2)});
\node[left] at(0,1){$1$};
\node[left] at(0,2){$2$};
\node[left] at(0,3){$3$};
\node[below] at(1,1){$q$};
\draw [xstep=0.1, ystep=0.2, gray, line width=0.2pt, dashed](0,1)grid(1,3);
\end{tikzpicture}
\end{center}
\subcaption{$\mathsf{E}_{p^*}(\xi \mid \iota = \I)$ as a function of $q$.}
\end{subfigure}
\end{figure}
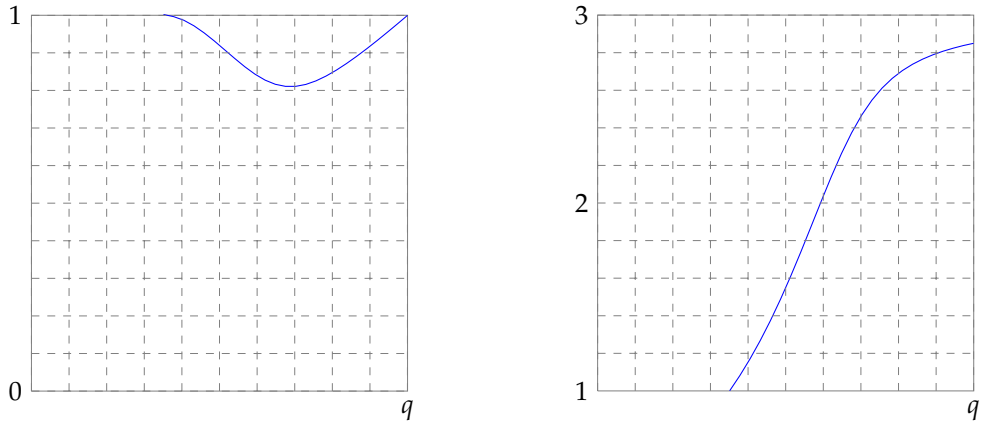

%\[\begin{aligned}
%p^*(\iota = \I) &= \frac{(1-k) - (1-k)(1-\beta)^n}{\beta}q_\I,&\quad&\mathsf{E}_{p^*}(\xi \mid \iota = \I) = \frac{n\beta~}{1 - (1-\beta)^n}& \\ p^*(\iota = \II) &= (1-k)\beta^{n-1}q_\II,&\quad&\mathsf{E}_{p^*}(\xi \mid \iota = \II) = 
%\end{aligned}\]
The expected number of children, conditional on the active player, are given by 
\[\mathsf{E}_{p^*}(\xi \mid \iota = \I) = \frac{n\beta}{1 - (1-\beta)^n}\text{ and }\mathsf{E}_{p^*}(\xi \mid \iota = \II) = n.\]
In fact, in the conditional game, the nodes controlled by Player II have exactly $n$ children almost surely. We have $\mathsf{E}_{p^*}(\xi \mid \iota = \I) \leq \mathsf{E}_{p^*}(\xi \mid \iota = \II)$. The unconditional expected number of children under $p^*$ is
\[\mathsf{E}_{p^*}(\xi) = q_\I(1-k)n + q_\II(1-k)n\beta^{n-1}.\]

Finally, under $p^*$ the capacity $\gamma$ is uniformly distributed on $[k,1]$, and the random variables $\xi$ and $\gamma$ are independent conditionally on $\iota$.
\end{exl}

\subsection{The proof of Theorem \ref{thm.upperconditional}}\label{secn.law_of_p*}
Let $V$ denote the event $\{\omega \in \Omega : v_{\omega} \geq k\}$. For a Borel set $B \subseteq \Omega$ let $B^* = \{\omega \in \Omega: \omega^* \in B\}$. We need to show that 
\begin{align*}
&\mathbb{P}(E(i,c,0)^* \mid V) = p_{i}^*(c,0),\\
&\mathbb{P}\Big(E(i,c,n)^* \cap \bigcap_{\ell = 1}^{n}\{\omega(\ell) \in B_{\ell}\}^* \mid V\Big) = p_i^*(c,n)\prod_{\ell = 1}^{n}\mathbb{P}(B_{\ell}^* \mid V),
\end{align*}
where $i \in \{\I,\II\}$, $c \geq k$, $n \geq 1$, and $B_1, \dots, B_n$ are Borel subsets of $\Omega$. 

For each $\omega \in \Omega$ 
\[\iota_{\omega^*}(\oslash) = \iota_{\omega}(\oslash),\quad \gamma_{\omega^*}(\oslash) = \gamma_{\omega}(\oslash),\text{ and }\quad \xi_{\omega^*}(\oslash) = \sum_{j = 1}^{\xi_{\omega}(\oslash)}1_{\{v_{\omega}(j) \geq k\}}.\]

We $E(i,c,0)^* \cap V = E(i,c,0)$. Indeed, if the value of the game is at least $k$, then either the node $\oslash$ has no children, or it has at least one child with a value of at least $k$. Hence
\[\mathbb{P}(E(i,c,0)^* \mid V) = \frac{\mathbb{P}(E(i,c,0))}{\mathbb{P}(V)} = \frac{p_{i}^*(c,0)}{\beta}.\]
This establishes the first of the two equations. We turn to the second one. We have:
\begin{align*}
E(i,c,n)^* &= \Big\{\iota_{\omega}(\oslash) = i,\,  \gamma_{\omega}(\oslash) \geq c,\, \sum_{j = 1}^{\xi_{\omega}(\oslash)}1_{\{v_{\omega}(j) \geq k\}} = n\Big\}\\ &= \bigcup_{m \geq n}\Big(E(i,c,m) \cap \Big\{\sum_{j = 1}^{m}1_{\{v_{\omega}(j) \geq k\}} = n\Big\}\Big)\\ &= \bigcup_{\substack{m \geq n\\1 \leq j_1 < \cdots < j_n \leq m}}\Big(E(i,c,m) \cap \bigcap_{\ell = 1}^{n}\{v_{\omega}(j_\ell) \geq k\} \bigcap_{\substack{j \in \{1,\ldots,m\}\\j \notin \{j_1, \ldots, j_n\}}}\{v_{\omega}(j) < k\}\Big).
\end{align*}

Take an $\omega \in E(i,c,n)^*$ and choose the indicies $m \geq n$ and $1 \leq j_1 < \cdots < j_n \leq m$ as in the union above, that is, so that for each $j \in \{1, \dots, m\}$ it holds that $v_{\omega}(j) \geq k$ if and only if $j \in \{j_1,\ldots,j_n\}$. Then $j_1, \dots, j_n$ index the $n$ level-1 subgames of $\omega$ having a value of at least $k$. For each $\ell \in \{1, \dots, n\}$ it holds that $(\omega)^*(\ell) = (\omega(j_{\ell}))^*$: that is to say, the $\ell$th subgame of the conditional game $\omega^*$ is the conditional game of the $j_{\ell}$th subgame of $\omega$. Consequently, $\{\omega(\ell) \in B_{\ell}\}^* = \{\omega(j_{\ell}) \in B_{\ell}^*\}$. Recalling also that $v_{\omega}(j) = v_{\omega(j)}(\oslash) = v_{\omega(j)}$ we obtain
\begin{align}
& E(i,c,n)^* \cap \{\omega(1) \in B_{1}\}^* \cap \cdots \cap \{\omega(n) \in B_{n}\}^*= \nonumber\\ &\bigcup_{\substack{m \geq n\\1 \leq j_1 < \cdots < j_n \leq m}}\Big(E(i,c,m) \cap \bigcap_{\ell = 1}^{n}\Big\{\begin{array}{c}v_{\omega(j_{\ell})} \geq k\text{ and}\\\omega(j_{\ell}) \in B_{\ell}^*\end{array}\Big\} \bigcap_{\substack{j \in \{1,\ldots,m\}\\j \notin \{j_1,\dots, j_n\}}}\{v_{\omega(j)} < k\}\Big).\label{eqn.union}
\end{align}

Using \eqref{eqn.measure2} we compute the probability of the event under the union in \eqref{eqn.union}:
\begin{align*}
& \mathbb{P}\Big(E(i,c,m) \cap \bigcap_{\ell = 1}^{n}\Big\{\begin{array}{c}v_{\omega(j_{\ell})} \geq k\text{ and}\\\omega(j_{\ell}) \in B_{\ell}^*\end{array}\Big\} \bigcap_{\substack{j \in \{1,\ldots,m\}\\j \notin \{j_1,\dots, j_n\}}}\{v_{\omega(j)} < k\}\Big) = \\ & p_i(c,m) \cdot \prod_{\ell = 1}^{n}\mathbb{P}\Big(\Big\{\begin{array}{c}v_{\omega} \geq k\text{ and}\\\omega \in B_{\ell}^*\end{array}\Big\}\Big) \prod_{\substack{j \in \{1,\ldots,m\}\\j \notin \{j_1,\dots, j_n\}}}\mathbb{P}(\{v_{\omega} < k\}) \\ & p_i(c,m) \beta^n \alpha^{m - n} \cdot \prod_{\ell = 1}^{n}\mathbb{P}(B_\ell^* \mid V).
\end{align*}

Let $i = \I$. Then $E(\I,c,n)^* \subseteq V$: indeed, since $n \geq 1$ and $c \geq k$, on the event $E(\I,c,n)^*$ the capacity at the root is at least $k$, and there is at least one level-1 subgame with a value of $k$ or higher. It follows that
\begin{align*}
& \mathbb{P}\Big(E(\I,c,n)^* \cap \bigcap_{\ell = 1}^{n}\{\omega(\ell) \in B_{\ell}\}^* \mid V\Big) = \\ & \frac{1}{\beta} \mathbb{P}\Big(E(\I,c,n)^* \cap \bigcap_{\ell = 1}^{n}\{\omega(\ell) \in B_{\ell}\}^*\Big) = \\ & \frac{1}{\beta}  \sum_{m \geq n} \binom{m}{n} p_\I(c,m) \beta^{n} \alpha^{m - n} \cdot \prod_{\ell = 1}^{n}\mathbb{P}(B_\ell^* \mid V) =\\ & p_\I^*(c,n) \cdot \prod_{\ell = 1}^{n}\mathbb{P}(B_\ell^* \mid V). 
\end{align*}

Let $i = \II$. Notice that $\xi_{\omega}(\oslash) = n$ everywhere on the event $E(\II,c,n)^* \cap V$. It follows that under the conditional measure $\mathbb{P}(\cdot \mid V)$ all of the events in the union \eqref{eqn.union} have probability $0$, except possibly the one corresponding to $m = n$. This latter event is in fact contained in $V$. Hence
\begin{align*}
& \mathbb{P}\Big(E(\II,c,n)^* \cap \bigcap_{\ell = 1}^{n}\{\omega(\ell) \in B_{\ell}\}^* \mid V\Big) = \\ & \frac{1}{\beta} p_\II(c,n) \beta^n \cdot \prod_{\ell = 1}^{n}\mathbb{P}(B_\ell^* \mid V) = \\ & p_\II^*(c,n) \cdot \prod_{\ell = 1}^{n}\mathbb{P}(B_\ell^* \mid V). 
\end{align*}

\section{Avoiding Player II's nodes}\label{secn.avoid}
In this section we consider a scenario under which Player I is prohibited from taking an action that would lead to Player II's node with more than one child. We think of this scenario as a proxy for the situation where Player I is reluctant to concede a turn of the game to her opponent. Such behaviour on the part of Player I effectively deprives Player II of any real choice in the game, rendering him a dummy: indeed, the only nodes assigned to him that could every be visited during play are end nodes or the nodes with a single child.

What payoff can Player I still guarantee (with positive probability) under this scenario?  
It is clear that the ``avoidance game" is harder for her to play. And yet, as is shown below, if Player I can guarantee a payoff of at least $k$ with positive probability in the absence of any restrictions on her behaviour, she can do so without ever visiting Player II's nodes with more than one child. 

Given a game $\omega$, define the \textit{avoidance game}\label{def.avoid} $\omega'$\label{def.omega'} by setting the capacity to zero at any Player II's node with more than one child: thus $T_{\omega'} = T_{\omega}$, $\iota_{\omega'} = \iota_{\omega}$, while $\gamma_{\omega'}$ is defined as follows: $\gamma_{\omega'}(h) = 0$ whenever $\iota_{\omega}(h) = \II$ and $\xi_{\omega}(h) \geq 2$, and $\gamma_{\omega'}(h) = \gamma_{\omega}(h)$ otherwise. Setting capacity at a node to zero effectively prohibits Player I from visiting that node. Clearly, the value $v_{\omega'}$ of the game $\omega'$ is not greater than the value of $\omega$.

Define a primitive distribution $p'$ by letting $p_{i}'(k,n)$ be $0$ if $i = \II$ and $n \geq 2$, and letting it be equal to $p_{i}(k,n)$ otherwise. We state the following lemmas for the sake of completeness. The proofs are easy and are omitted.

\begin{lemma}
The map $\omega \mapsto \omega'$, $\Omega \to \Omega$ is Borel measurable.
\end{lemma}

\begin{lemma}\label{lem.p'}
The distribution of the game $\omega'$ is given by the measure $\mathbb{P}_{p'}$: that is to say $\mathbb{P}_{p}(\{\omega' \in B \}) = \mathbb{P}_{p'}(B)$ for each Borel set $B \subseteq \Omega$.
\end{lemma}

The main message of this section is the following rather surprising result (it follows from Lemma \ref{lem.p'} and Theorem \ref{thm.prob>0}).    
\begin{cor}\label{thm.phasetransitions}
Let $k > 0$ and suppose that $p(\{\gamma < k\}) > 0$. Then $\mathbb{P}_{p}(\{v_{\omega} \geq k\}) > 0$ if and only if $\mathbb{P}_{p}(\{v_{\omega'} \geq k\}) > 0$.
\end{cor}
%Combining Corollaries \ref{thm.phasetransitions} and \ref{thm.prob>0escape} we obtain the following conclusion.

\begin{cor}
Suppose that $p$ is an escape model. Let $k > 0$. Then $\mathbb{P}_{p}(\{v_{\omega} \geq k\}) > 0$ if and only if $\mathbb{P}_{p}(\{v_{\omega'} \geq k\}) > 0$ if and only if $d(k) > 1$.
\end{cor}

\begin{exl}\rm
Consider the setup of Example \ref{exl.fraclin}. The vgf corresponding to the measure $p'$ (where we take $k = 1$) is given 
\[f_{1}'(x) = 1 - l(1-x)\Big(q_\I\frac{1}{1 - l x} + q_\II(1 - l)\Big).\]
Computing the smallest fixed point of the function, we obtain
\[\mathbb{P}(v_{\omega'} = 0) = \frac{1-l+l^{2}-l^{2}q}{l(1-l+l^{2}+l(1-l)q)}.\]
The probabilities $\mathbb{P}(v_{\omega'} = 0)$ and $\mathbb{P}(v_{\omega} = 0)$ are pictured in Figure \ref{fig.nonadv} (the latter is the same as the red line in Figure \ref{fig.fraclin}).

\begin{figure}
\caption{}
\label{fig.nonadv}
\begin{subfigure}{0.5\textwidth}
\begin{center}
\begin{tikzpicture}[domain=0:1,scale=5]
\draw[very thin,color=gray](0,0)--(1,0)--(1,1)--(0,1)--(0,0);
\draw[color=blue](0,1)--(0.1,1);
\node[below] at(1,0){$q$};
\draw[color=blue, domain=0.1:1] plot (\x,{(1-0.9+0.9^2-0.9^2*\x)/(0.9*(1-0.9+0.9^2+0.9*(1-0.9)*\x))});
\draw[color=red](0,1)--(0.1,1);
\draw[color=red, domain=0.1:1] plot (\x,{0.5*(2-0.9)*(1-\x) + 0.5*(4*(0.1/0.9)^2 + ((2-0.9)*(1-\x))^2)^(0.5)});
\draw [step=0.1, gray, line width=0.2pt, dashed](0,0)grid(1,1);
\end{tikzpicture}
\caption{The probabilities of the events $\{v_{\omega'} = 0\}$ (blue) and $\{v_{\omega} = 0\}$ (red) in Example \ref{exl.fraclin} as functions of $q$ for $l = 0.9$.}
\end{center}
\end{subfigure}\hspace{1cm}
\begin{subfigure}{0.5\textwidth}
\begin{center}
\begin{tikzpicture}[domain=0:1,scale=5]
\draw[very thin,color=gray](0,0)--(1,0)--(1,1)--(0,1)--(0,0);
\draw[color=red, domain=0.3508:1] plot (\x,{0.5*(2 - 3*\x) + 0.5*((2 - 3*\x)^2 + 4*(1/19))^(0.5)});
\draw[color=blue, domain=0.3508:1] plot (\x,{-0.5 + (-0.75+1/(0.95*\x))^(0.5)});
\node[below] at(1,0){$q$};
\draw [step=0.1, gray, line width=0.2pt, dashed](0,0)grid(1,1);
\end{tikzpicture}
\caption{The probabilities of the events $\{v_{\omega'} < 0.05\}$ (blue) and $\{v_{\omega} < 0.05\}$ (red) in Example \ref{exl.fraclin} with $n = 3$ for $k = 0.05$.}
\end{center}
\end{subfigure}
\end{figure}

Consider Example \ref{exl.tree}. Note that, since all nodes have $n$ children, in $\omega'$ all the nodes controlled by Player II have capacity zero. The vgf corresponding to $p'$ is 
\[f_{k}'(x) = 1 - (1-k)q + (1-k)qx^n,\]
and its smallest fixed point, for the ternary case $n = 3$ is
\[\mathbb{P}(v_{\omega'} < k) = -\tfrac{1}{2} + \sqrt{\frac{1}{(1-k)q} - \tfrac{3}{4}}.\]
This probability, along with $\mathbb{P}(v_{\omega} < k)$ is pictured in the right panel of Figure \ref{fig.nonadv} for $k = 0.05$ as a function of $q$ (the latter graph is the same as that in Figure \ref{fig.tree}).
\end{exl}

We now take a closer look at the set of $k$-optimal strategies in $\omega'$ and compare them to those in $\omega$. For clarity we assume $p$ to be an escape model (in which case $p'$ is as well). 

Fix a $k > 0$. One can think of player I's $k$-optimal strategy in $\omega'$ as simply an infinite play in the tree $T_{\omega}$ that never visits a node with a capacity less than $k$, nor a node where Player II has more than one child. Clearly, a $k$-optimal strategy in $\omega'$ is also a $k$-optimal strategy in $\omega$, but, in general, Player I has $k$-optimal strategies in $\omega$ that are unavailable to her in $\omega'$.

Recall that the tree of the $k$-conditional game characterizes Player I's $k$-optimal strategies. We can thus get a sense of the relative ``sizes" of the sets of $k$-optimal strategies in $\omega$ and in $\omega'$ by comparing the game trees of the respective $k$-conditional games. 

Thus suppose that $v_{\omega'} \geq k$ and consider $T_{\omega}^{*}$ and $T_{\omega'}^{*}$: the former is the tree of the conditional game associated with $\omega$, and the latter is the tree of the conditional game associated to $\omega'$ (both prior to reordering). The trees satisfy the inclusion $T_{\omega^\prime}^{*} \subseteq T_{\omega}^{*}$ as can be seen from the fact that $v_{\omega'}(h) \leq v_{\omega}(h)$ for each node $h \in T_{\omega'}$. Since $p$ and $p'$ are escape models, the trees $T_{\omega}^{*}$ and $T_{\omega'}^{*}$ contain no end nodes, and are thus completely described by their respective boundaries. The boundary of $T_{\omega'}^{*}$ is easy to describe explicitly: it is the set of infinite plays $p \in \partial T_{\omega}$ that do not pass through the nodes with a capacity less than $k$, nor through Player II's nodes with more than one child. 

Applying Corollary \ref{thm.GW*} to the probability measure $p'$ and using Lemma \ref{lem.p'} we obtain the following statement.
\begin{cor}\label{thm.GW'}
Suppose that $\mathbb{P}_{p}(\{v_{\omega'} \geq k\}) > 0$. Conditional on the event $\{\omega \in \Omega:v_{\omega'} \geq k\}$, the distribution of the tree $T_{\omega'}^*$ is the Galton-Watson measure generated by an offspring distribution with mean $d(k)$. 
\end{cor}

This result leads to an interpretation of $d(k)$ as the average number of $k$-optimal actions Player I has in the avoidance game. Or equivalently, as the average number of  actions that allow her to guarantee a payoff of at least $k$ while at the same time avoid visiting nodes where her opponent has more than one move. 

We can compare the two trees (and their respective boundaries) with the help of branching numbers (Lyons and Peres \cite[\S 1.8]{LyonsPeres}).
\begin{cor}
Suppose that $p$ is an escape model. Let $k > 0$ and suppose that $d(k) > 1$. Then, $\mathbb{P}_{p}$-almost surely on the event $\{\omega \in \Omega: v_{\omega'} \geq k\}$:
\begin{itemize}
\item[(i)] The branching number of $T_{\omega}^{*}$ equals $\mathsf{E}_{p^*}(\xi)$.
\item[(ii)] The branching number of $T_{\omega'}^{*}$ equals $d(k)$. 
\item[(iii)] If, moreover, $p(\{\iota = \II\} \cap \{\gamma \geq k\} \cap \{\xi \geq 2\}) > 0$, then $\partial T_{\omega'}^{*}$ is a proper subset of $\partial T_{\omega}^*$.
\end{itemize}
\end{cor} 
\begin{proof}
The corollary follows from the following well-known fact (see e.g. Lyons and Peres \cite{LyonsPeres}[\S 1.8 and Corollary 5.10]): consider a Galton-Watson measure on trees generated by the offspring distribution with a mean greater than 1. Almost surely on the event of non--extinction of the tree, its branching number equals the mean of the offspring distribution. 

To obtain claim (i), we apply this fact to the distribution of the tree $T_{\omega}^{*}$, using Corollary \ref{thm.GW*}, and noting that the event $\{v_{\omega'} \geq k\}$ is a subset of $\{v_{\omega} \geq k\}$, and that $\mathsf{E}_{p^*}(\xi) > 1$. To obtain (ii), we apply the above fact to the distribution of the tree $T_{\omega'}^{*}$ using Corollary \ref{thm.GW'}. To obtain claim (iii) use equation \eqref{eqn.Ep*(xi)}: 
\[\mathsf{E}_{p^*}(\xi) - d(k) = \mathsf{E}_{p}(1_{\{\iota = \II\} \cap \{\gamma \geq k\} \cap \{\xi \geq 2\}} \xi\beta^{\xi-1}),\]
where $\beta = \mathbb{P}_{p}(\{v_{\omega} \geq k\})$. Since $d(k) > 1$, we have $\beta > 0$ and hence $\mathsf{E}_{p^*}(\xi) > d(k)$.
\end{proof}

\section{Discussion and open questions}
\subsection{On Player II's $k$-optimal strategies}
We have analysed the set of Player I's $k$-optimal strategies with the help of the conditional game. Here we would like to reflect on the difference in the nature of $k$-optimal strategies of Player I and those of Player II. 

Consider the following Player I's strategy for a game with value $k$: always choose the youngest (i.e. the one with the lowest index) child of the current node having a value of at least $k$. This strategy (let us call it ``simple") is $k$-optimal. The counterpart of the simple strategy for Player II -- always choosing the youngest child with a value no greater than $k$ -- need not be $k$-optimal. For instance, take a game where Player II is the only player, each node has 2 children, the younger child (the one with the index $1$) has capacity $1$, while the older child the capacity $0$. The simple strategy is not $0$-optimal. Any $0$-optimal strategy for Player II must choose the older child at some point.  

A question arises whether examples like this one are exceptional in the measure-theoretic sense. Would not the simple strategy be $k$-optimal in all games with a value of at most $k$ apart from a set of measure zero? In general the answer is no. 

Consider Example \ref{exl.fraclin} with $l = 0.9$ and $q = 0.1$, which is slightly below the critical level (of $0.1021$), so that $\alpha = 1$. Thus, Player II does have a $0$-optimal strategy with probability 1. Consider what happens if Player II uses the simple srtategy instead. Since all nodes of the game tree will have value $0$ with probability $1$, the simple strategy boils down to choosing the first child. 

Let $S_{\omega} \subseteq T_{\omega}$ denote the tree consisting of the nodes that are consistent with Player II's simple strategy. The tree $S_{\omega}$ is distributed according to a Galton-Watson measure with the mean $0.1 \cdot 9 + 0.9 \cdot 0.9 \cdot 1 = 1.71$ (since a node controlled by Player I has $9$ children on average, while a node controlled by Player II has no children or exactly $1$ child, depending on whether in $T_{\omega}$ the node had any children). Thus $S_{\omega}$ has an infinite branch with positive probability. And hence Player I has a positive probability of winning against Player II's simple strategy. We conclude that, with positive probability, Player II's simple strategy is not $0$-optimal.

This illustrates that Player II's $k$-optimal strategies require a different approach than that we have employed in Section \ref{secn.conditional}. Finding a suitable approach is an interesting direction for future work.  

\subsection{On finite games}\label{subsecn.finite}
Our model encompasses certain classes of finite games. To illustrate, suppose that the mean of the offspring distribution $\mathsf{E}_{p}(\xi)$ is smaller than $1$, that the capacity equals $1$ whenever $\xi > 0$, and that it is uniformly distributed on $[0,1]$ if $\xi = 0$. Under these assumptions the game tree $T_{\omega}$ is finite almost surely, and the payoff equals the capacity at the end node. Thus, we have a model of finite games with payoffs randomly and independently assigned to the end nodes of the game tree. 

The model outlined above resembles that in Arieli and Babichenko \cite{Arieli} with the difference that our game tree is random; in particular, there is a positive probability for the root of the game tree to be its only node. Thus the two models are distinct. Developing a framework to encompass both these models as special cases might be a fruitful avenue for future research.

\subsection{Subgame perfect equilibrium in multiplayer perfect information games}\label{subsecn.multiplayer}
Multiplayer perfect information games with semicontinuous payoffs have been the subject of much work (Flesch, Kuipers, Mashiah-Yaakovi, Schoenmakers, Solan, Vrieze  \cite{FleschKuipers}, Purves and Sudderth \cite{PurvesSudderth}, Flesch and Predtetchinski \cite{FleschPred}, Flesch, Herings, Maes, and Predtetchinski \cite{FleschHerings}). Nevertheless, we believe that the modelling technique of this paper, suitably adapted, could offer a new perspective on the topic. Here we suggest one particular question. 

Consider perfect information games played by an infinite sequence of players $0, 1, 2, 
\ldots$, player $t \in \N$ moving only once, in period $t$. Suppose that each has a lower semicontinuous payoff. It is known that not all games of this class admit a subgame perfect $\epsilon$-equilibrium. But how ``large" is the set of games that do have one? One possible approach to this question is probabilistic: suppose that each player's payoff function is generated randomly, by independently assigning the capacities to the nodes, as is done in this work. What is the measure of games that have a subgame perfect $\epsilon$-equilibrium?

\subsection{On the non-adversarial case}    
Non-adversarial case is the special case of the model with $q_\I = 1$, so that all the nodes of the game tree are assigned to Player I. We have mentioned the non-adversarial case when discussing the examples in Section \ref{secn.examples}; in both examples, the non-adversarial case boils down to the question of (non-)extinction of a certain branching process. This observation can be generalized to escape models. 

Consider an escape model with $q_\I = 1$, and let $T_{\omega}^k$ be the subtree of $T_{\omega}$ consisting of the nodes with a capacity of at least $k$. Then the value of $\omega$ is at least $k$ precisely when $T_{\omega}^k$ has an infinite branch. The tree $T_{\omega}^k$ is governed by a Galton-Watson measure, and the corresponding offspring distribution is that of the random variable $1_{\{\gamma \geq k\}}\xi$. In particular, $\{v_{\omega} \geq k\}$ has a positive probability if and only if $\mathsf{E}_{p}(1_{\{\gamma \geq k\}}\xi) > 1$ or $p(\{\gamma \geq k\} \cap \{\xi = 1\}) = 1$.

In general, the non-adversarial case is not entirely trivial. For example, if the capacity is independent of the number of children, Player I might strive to finish the game by reaching an end node, and do so sooner rather than later, lest she encounters a node with a low capacity. Consider, for example, the non-adversarial case of the model described in Subsection \ref{subsecn.finite}. The value is then the maximum of a random number $\eta$ of independent uniformly distributed random variables, where the random variable $\eta$ is the number of end nodes in the family tree of a subcritical branching process. See Nariyuki \cite{Nariyuki} for the properties of its distribution.

\end{document}